\documentclass[acmsmall]{acmart}

\usepackage{multirow}
\usepackage{enumitem}
\usepackage{xcolor}
\usepackage{bbding}

\begin{document}

\title{Cascading Residual Graph Convolutional Network for Multi-Behavior Recommendation}

\author{Mingshi Yan}
\affiliation{%
  \institution{Qilu University of Technology (Shandong Academy of Sciences) \& Dalian Minzu University}
  \country{China}
  \postcode{116600}
}
\email{neo.ms.yan@gmail.com}

\author{Zhiyong Cheng$^{\dag}$}
\affiliation{%
  \institution{Qilu University of Technology (Shandong Academy of Sciences)}
  \country{China}
  \postcode{250014}
}
\email{jason.zy.cheng@gmail.com}

\author{Chen Gao}
\affiliation{%
  \institution{Tsinghua University}
  \country{China}
  \postcode{100084}
}
\email{chgao96@gmail.com}

\author{Jing Sun}
\affiliation{%
  \institution{Dalian Minzu University}
  \country{China}
  \postcode{116600}
}
\email{jingsun@dlnu.edu.cn}

\author{Fan Liu}
\affiliation{%
  \institution{National University of Singapore}
  \postcode{117417}
    \country{Singapore}
  }
 \email{liufancs@gmail.com}

\author{Fuming Sun}
\affiliation{%
  \institution{Dalian Minzu University}
  \country{China}
  \postcode{116600}
}
\email{sunfuming@dlnu.edu.cn}

\author{Haojie Li}
\affiliation{%
  \institution{Dalian University of Technology}
  \country{China}
  \postcode{116600}
}
\email{hjli@dlut.edu.cn}

\thanks{$^{\dag}$ Corresponding Author. \\ This work was finished when Mingshi Yan was a visting student at the Shandong Artificial Intelligence Institute, Qilu University of Technology (Shandong Academy of Sciences), under the supervision of Dr. Zhiyong Cheng.
}


\begin{abstract}
Multi-behavior recommendation exploits multiple types of user-item interactions, such as \emph{view} and \emph{cart}, to learn user preferences and has demonstrated to be an effective solution to alleviate the data sparsity problem faced by the traditional models that often utilize only one type of interaction for recommendation. In real scenarios, users often take a sequence of actions to interact with an item, in order to get more information about the item and thus accurately evaluate whether an item fits their personal preferences. Those interaction behaviors often obey a certain order, and more importantly, different behaviors reveal different information or aspects of user preferences towards the target item. Most existing multi-behavior recommendation methods take the strategy to first extract information from different behaviors separately and then fuse them for final prediction. However, they have not exploited the connections between different behaviors to learn user preferences. Besides, they often introduce complex model structures and more parameters to model multiple behaviors, largely increasing the space and time complexity. 
In this work, we propose a lightweight multi-behavior recommendation model named \emph{\textbf{C}ascading \textbf{R}esidual \textbf{G}raph \textbf{C}onvolutional \textbf{N}etwork}  (CRGCN for short) for multi-behavior recommendation, which can explicitly exploit the connections between different behaviors into the embedding learning process without introducing any additional parameters (with comparison to the single-behavior based recommendation model). In particular, we design a cascading residual graph convolutional network (GCN) structure, which enables our model to learn user preferences by continuously refining the embeddings across different types of behaviors. The multi-task learning method is adopted to jointly optimize our model based on different behaviors. Extensive experimental results on three real-world benchmark datasets show that CRGCN can substantially outperform the state-of-the-art methods, achieving 24.76\%, 27.28\%, and 25.10\% relative gains on average in terms of HR@K (K=$\{10,20,50,80\}$) over the best baseline across the three datasets. Further studies also analyze the effects of leveraging multi-behaviors in different numbers and orders on the final performance. 

\end{abstract}





\maketitle

\section{Introduction}

Personalized recommender systems, which find information and products that are of interest or need for, achieve a great success in the information-overload era~\cite{zhang2019deep}. Collaborative Filtering (CF)~\cite{SarwarKKR01, KorenBV09, Cheng2019mmalfm}, which learns user preferences from user-item interaction data, is one of the most successful and widely-used models in recommender systems.  Along with the development of recommendation techniques, plenty of CF models~\cite{NingK11, Koren08, RendleFGS09} have been proposed, from matrix factorization methods~\cite{KorenBV09,PMF} to deep neural network models~\cite{HeLZNHC17}, and to the recent advances of graph neural network models~\cite{Wang0WFC19, LightGCN}. Most CF models only consider one type of behavior (such as the \emph{buy} behavior on e-commerce platforms). With the huge number of products available for selection on e-commerce platforms, the number of products ultimately purchased by a user is very small, leading to extremely sparse data on \emph{buy} behavior. As a result, the CF models relying on a single-type behavior cannot well capture user preferences with limited interaction data (\textit{i.e.}, data sparsity or cold-start problem)~\cite{LightGCN}, resulting in dramatic performance degradation. Fortunately, when interacting with information systems, there are other behaviors, such as \emph{view} and \emph{collect}, which also provide interaction information between users and items. This motivates studies on leveraging multi-behavior information to assist in learning user preferences and thus alleviates the data sparsity issue. The technique, which exploits multiple types of behavior information for recommendation, is also called \textit{multi-behavior recommendation}~\cite{QiuLGSZN18, LeeHHRK15, ZhangMCX20, XiaHXDLB21, XiaHXDZB20, SchlichtkrullKB18, JinG0JL20, GaoHGCFLCYSJ21}. 

Among existing multi-behavior recommendation methods, many methods treat other types of behavior data (besides the target behavior) as auxiliary data to help learn user preferences in the training process. For example, the early matrix factorization-based model CMF~\cite{CMF} separately performs matrix factorization on multiple matrices constructed from data of different behaviors to learn shared user and item representations. Recent GCN-based model MBGCN~\cite{JinG0JL20} learns user embeddings by propagating item nodes' embeddings based on different user-item behavior propagation layers; and the final user embeddings are obtained by aggregating the embeddings learned from different user-item behaviors according to their contributions. The limitation of those methods is that they have not exploited the information of user preferences contained in the connections between different behaviors that often happen in a certain order in real scenarios, \emph{e.g., view->cart->buy}. In fact, the sequential behaviors of users interacting with items often disclose different levels of user preferences toward the target item. For example, when a user is attracted to an item by its easily-observable features, such as appearance or brand, she will click the item and take a look at more information about the item (\emph{view} behavior);  if the user is still interested in the item after getting more information about the item, she will put it into the shopping cart (\emph{cart} behavior); the user will finally purchase the item if the item satisfies her after carefully examining all aspects of the item or comparing it with other candidates (\emph{buy} behavior). The behaviors at different positions in the sequence reveal user preferences to items at different levels. The preference dependence information contained in the behavior sequence is beneficial to user preference modeling. As far as we know, only a few methods in the literature attempt to model~\cite{GaoHGCFLCYSJ21, LoniPLH16}  the dependent relationship between behaviors to exploit multiple types of behaviors. A typical example is NMTR~\cite{GaoHGCFLCYSJ21}, which correlates the model predictions of each behavior type in a cascading manner and develops multiple neural network modules to learn user preferences from the shared embedding layer under different behaviors. 

Despite the progress, there are still two major limitations in existing works as follows. Firstly, \textbf{the preference information conveyed in the behavior sequence has not been well exploited in user preference modeling.} Existing works of multi-behavior recommendation often independently model the different types of behaviors, which have not well exploited the preference dependence among different behaviors. Although a few recent works, such as NMTR, have considered the connection between different behaviors, it has not explicitly incorporated the connection information into the embedding learning process. Instead, it exploits the connection information in the predicted scores to guide the embedding learning based on different behaviors. Besides, \textbf{multi-behavior recommendation models are often implemented at the cost of complex model structures and high computational costs.}
Existing multi-behavior recommendation models generally ignore the model complexity when pursuing more accurate recommendation performance. In fact, the common paradigm of modeling each type of behavior separately in existing methods will introduce a large number of additional trainable parameters. Moreover, to achieve higher recommendation accuracy, they often adopt some advanced techniques in the model, such as attention mechanisms, which often further increase the computational complexity of the model. In real applications, efficiency is also an important aspect of recommender systems. A model with comparable recommendation accuracy but fewer parameters and lower computational complexity will be more desirable.

Motivated by the above considerations, we propose a novel lightweight graph neural network-based model named \emph{\textbf{C}ascading \textbf{R}esidual \textbf{G}raph \textbf{C}onvolutional \textbf{N}etwork} (CRGCN), which leverages the relations among the sequential behaviors to gradually learn and refine user preference representations. We deem that the sequence of behaviors corresponds to the user's decision-making process, \textit{e.g.}, \emph{view, collect}, and \emph{buy}. The different types of behaviors in a sequence reflect user's preferences to items at different levels. For example, a \emph{buy} behavior denotes that the item can well satisfy the user's preference, and a \emph{view} behavior indicates that the user is interested in the item in some aspect, such as appearance. With this consideration, we propose to model user preferences (\textit{i.e.}, user embedding in the model) by continuously refining it across the sequence of behaviors. For ease of presentation, we define the initialized embedding as the \textbf{\textit{basic features}}, and the features learned from each behavior as the \textbf{\textit{behavioral features}}. More precisely, we first build cascading sequences based on the user's decision process, from the initial \emph{view} behavior to the final \emph{buy} behavior. For each type of behavior, a residual block is assigned to learn the behavioral features based on \emph{the output of the previous block} and \emph{the interaction data of this behavior}, and then output a fusion of \emph{the learned features} and \emph{the output of the previous block from a short-cut connection} (see Fig.~\ref{fig:global} for the model structure). The short-cut connection is to preserve the information learned from the previous type of behavior. The behavioral features in each block are learned by the LightGCN model~\cite{LightGCN}, due to its impressive performance and lightweight design by removing the feature transformation and non-linear operation. The cascading structure is adopted to connect multiple residual blocks according to the occurrence sequence of behaviors. User preferences can be continuously learned and refined through the sequence of behaviors. In this way, our model explicitly exploits the preference relation between different behaviors in the embedding learning process. Finally, we utilize the multi-task learning approach, which can effectively exploit all the data simultaneously to jointly optimize multiple behaviors to learn the user/item embeddings. It is worth mentioning that our model does not introduce any trainable parameters besides the ones used for initializing the user and item embeddings. Therefore, our model enjoys the advantage of low complexity and is easily trained. We conduct extensive experiments and ablation studies on three real-world benchmark datasets, Tmall, Beibei, and Jdata, to evaluate both the effectiveness and efficiency of our proposed CRGCN model. The evaluation results demonstrate that the proposed CRGCN significantly outperforms the state-of-the-art recommendations, including single-behavior models~\cite{RendleFGS09, LightGCN} and multi-behavior methods~\cite{SchlichtkrullKB18, GaoHGCFLCYSJ21, JinG0JL20, XiaHXDLB21}. Remarkably, it achieves 24.76\%, 27.28\%, and 25.10\% relative gains on average in terms of HR@K (K=$\{10,20,50,80\}$) over the best baseline across the three datasets, respectively. Interestingly, we observed that the one-layer (for each residual block) can already achieve very impressive performance in experiments. Furthermore, CRGCN achieves the best performance in handling cold-start users even without user-item interactions from auxiliary behaviors in the training dataset.

To summarize, the main contributions of this work are as follows:
\begin{itemize}
  
  
  \item We highlight the importance of modeling the relations among multi-behaviors into user modeling in recommendation, and propose to directly exploit such relations into the embedding learning process by updating user preference along with the behaviors in an order. As far as we know, this is the first work to  take the cascading relations among multi-behaviors directly into the embedding learning for recommendation.
  
  \item We propose a multi-behavior recommendation model called CRGCN, which consists of a set of LightGCNs (corresponding to the sequence of behaviors) with a residual design, aiming to preserve the features learned from the previous behavior to the next behavioral feature learning. Our model is lightweight and does not introduce any additional trainable parameters into the model with comparison to the standard matrix factorization for single-behavior modeling. This makes our model enjoy a big advantage in space and time complexity.
  
  \item We conduct extensive experiments on three real-world datasets to verify the effectiveness of our CRGCN model. Experimental results show that CRGCN can achieve a remarkable improvement over the state-of-the-art models in terms of both accuracy and efficiency.
\end{itemize}

The remainder of this paper is structured as follows. Section~\ref{Related Work} briefly introduces related work, and Section~\ref{methodology} describes our CRGCN model in detail. Next, Section~\ref{experiment} introduces the experimental setup and reports the experimental results. Finally, Section~\ref{conclusion} concludes this paper.
\section{Related Work} \label{Related Work}
\subsection{Multi-Behavior Recommendation}
Multi-behavior recommendation refers to leveraging multiple types of user-item interactions to enhance recommendation performance~\cite{RioGNN, XuZY21}. 
Generally, it can improve the prediction ability of target behavior by extracting useful prediction signals from other types of behaviors. The existing works can be divided into two categories: shallow models and deep models based on neural networks.

The former methods are based on traditional machine learning techniques, \textit{i.e.}, shallow models. Earlier works exploited multi-behavior data with matrix factorization techniques~\cite{KorenBV09}. For example, Ajit \textit{et al.}~\cite{CMF} proposed a collective matrix factorization model (CMF) to decompose multiple matrices simultaneously with entity parameter sharing. Zhao~\textit{et al.}~\cite{ZhaoCHC15}  further extended CMF to perform matrix factorization of different behaviors by sharing items. 
In addition, some works exploited multiple behaviors by designing new sampling strategies~\cite{LoniPLH16, DingY0QLCJY18, GuoQTLMW17, QiuLGSZN18}.
Loni~\textit{et al.}~\cite{LoniPLH16} extended BPR~\cite{RendleFGS09} by designing a negative sampling strategy to sample the interaction data of user-item with different behaviors. 
Ding~\textit{et al.}~\cite{DingY0QLCJY18} further developed on this basis and improved the negative sampling strategy to make better use of the data. 
Guo~~\textit{et al.}~\cite{GuoQTLMW17} proposed to generate samples from multiple auxiliary behaviors according to the item-item similarity for training. 
Qiu~~\textit{et al.}~\cite{QiuLGSZN18} proposed an adaptive sampling strategy to solve the unbalanced correlation among different behaviors. Several studies consider temporal information in multi-behavior recommendations~\cite{MoeF04, LeeHHRK15, DongJ19}. 
The major limitation of these methods is the lack of exploration of the relationship among behaviors.

The latter ones are based on deep learning models.
In recent years, deep learning has exhibited a strong ability for representation learning~\cite{ChenYZWCN20, LuoCNYHZ18}, and has also been applied in multi-behavior recommender tasks. 
Gao~\textit{et al.}~\cite{GaoHGCFLCYSJ21} constructed multiple neural collaborative filtering (NCF) units to capture the complicated and multi-type interaction under each type of behavior. 
They consider the cascading relationship among different types of behaviors shown in user multi-behavior data, by re-using the previous-behaviors' prediction scores in the given behavior.
Jin~\textit{et al.}~\cite{JinG0JL20} constructed a unified graph to represent multi-behavior data, and learned the influence strength (to the target behavior) and semantics of different behaviors by user-item and item-item propagation layers, respectively. 
Xia~\textit{et al.}~\cite{XiaXHDB21} proposed a multi-behavior recommender framework with a graph meta-network, which incorporates the multi-behavior pattern modeling into a meta-learning paradigm for exploring the complex dependencies across different types of user-item interactions. 
Although great progress has been achieved by these methods, they have not fully considered the fine-grained relationships between behaviors, as discussed in the introduction. 
In addition, these methods often rely on heavy network structure and introduce more parameters to model the multi-behaviors, thus largely increasing the computational complexity.

In this work, we model different behaviors in the form of cascading residual blocks by effectively exploring the preference information connections between behaviors. We propose to refine the user embedding based on cascading residual blocks, corresponding to the user-decision process.
In addition, our model does not introduce any learnable parameters besides user/item embeddings, which enjoys the benefits of low computational complexity.

\subsection{Graph Convolutional Network for Recommendation} 

Graph Convolution Network (GCN)-based models have achieved outstanding success in a variety of applications~\cite{KipfW17}. The basic idea of GCN is to update a target node's embedding by iteratively aggregating information from its local graph neighbors. Due to its strong capability of representation learned from non-Euclidean structures, GCN has also been widely applied in recommender systems~\cite{QiuHLY20, 0002DWWTFC022, XieZHDN22}, since relations between users and items can be naturally represented by graph structures.

As for collaborative filtering, which is the most fundamental recommendation technique,
GCN-based models have shown strong performance~\cite{Wang0WFC19, LightGCN, WangJZ0XC20, ChenWHZW20,liu2022aagcn,liu2021impgcn}.
Wang \textit{et al.}~\cite{Wang0WFC19} modeled the higher-order connectivity information and recursively propagated embeddings non-linearly on the graph. He \textit{et al.}~\cite{LightGCN} deeply analyzed the effect of feature transformation and nonlinear activation.
Thus, the authors further proposed to replace nonlinear propagation with linear propagation, and retain only the most basic neighborhood aggregation components. Such an operation not only simplified the model structure but also made the model easier to implement and train. Wang \textit{et al.}~\cite{WangJZ0XC20} regarded the user-item interaction graph as an entanglement model of heterogeneous information and obtained the representation of the user's different intentions through embedding disentanglement. Moreover, GCN-based models are also widely deployed in multi-behavior recommendation tasks~\cite{XiaXHDB21, JinG0JL20, ZhangMCX20, XiaHXDLB21}. 
For example, Jin \textit{et al.}~\cite{JinG0JL20} built multi-behavior data into a unified heterogeneous graph, and then used GCN to learn the behavioral strength and user preferences. Xia \textit{et al.}~\cite{XiaHXDLB21} captured type-aware behavior collaborative signals through message propagation on heterogeneous graphs. Zhang \textit{et al.}~\cite{ZhangMCX20} modeled different graph networks for various behaviors to explore the commonality and specificity of user preferences in different behaviors, in which the GCN model helps improve the accuracy of feature extraction.

In our work, we propose cascading residual blocks based on graph convolutional networks, which can not only effectively extract the preference signal of each type of behavior but also refine user embedding by extracting useful information from the signal learned in each behavior across the residual blocks corresponding to the sequence of behaviors.

\section{methodology} \label{methodology}
\subsection{Problem Formulation}
In the real-world scenario of recommendation platforms, there are multiple types of interactions between users and items. 
However, traditional methods often only consider the user-item interactions under one specific behavior, \textit{i.e.}, target behavior~\cite{JinG0JL20,GaoHGCFLCYSJ21}. 
Other kinds of behaviors, such as \emph{view, cart, collect, etc.}, have not been well exploited. These auxiliary behavioral data provide rich information about user preferences and can be leveraged to model user preferences better, and thus can help alleviate data sparsity and cold-start problems, as well as improve recommendation performance. In this work, we propose a novel model to explicitly take the connections between different behaviors into the embedding learning process. Before formally introducing our model, we first introduce the key notations and problem setting.

Let $\mathcal{U}=\{u_{1}, \cdots, u_{m}, \cdots, u_{M}\}$ and $\mathcal{I}=\{i_{1}, \cdots, i_{n}, \cdots, i_{N}\}$ respectively be the set of users and items, in which $M$ and $N$ denote the number of users and items. We use $\mathcal{W}=[\boldsymbol{W}^{1}, \cdots, \boldsymbol{W}^{b}, \cdots, \boldsymbol{W}^{B}]$ to denote the list of interaction matrices sorted by a defined order, where $\boldsymbol{W}^{b}$ is the interaction matrix of the \textit{b-th} behavior and $\boldsymbol{W}^{B}$ is the target behavior. Specifically, interaction matrix $\boldsymbol{W}^{b}$ is binary, that is each entry in it has a value of 1 or 0, defined as follows:
\begin{equation}
  w^{b}_{ui} =
    \begin{cases}
        1,  & \text{if user $u$ has interacted with item $i$ under behavior $b$;} \\
        0, & \text{otherwise.} \\
    \end{cases}
\end{equation}

The studied problem is formulated as follows:

\noindent \textbf{Input:}  user set $\mathcal{U}$, item set $\mathcal{I}$, and the interaction matrix list $\mathcal{W}$.

\noindent \textbf{Output:} predicting a similarity score, which indicates the possibility that a user $u$ will take a target behavior (\textit{e.g.}, \emph{buy}) to an item $i$. The recommendation list can be generated by sorting items based on the similarity score in descending order.

\begin{figure}[htb]
  \centering
  \includegraphics[width=0.8\textwidth]{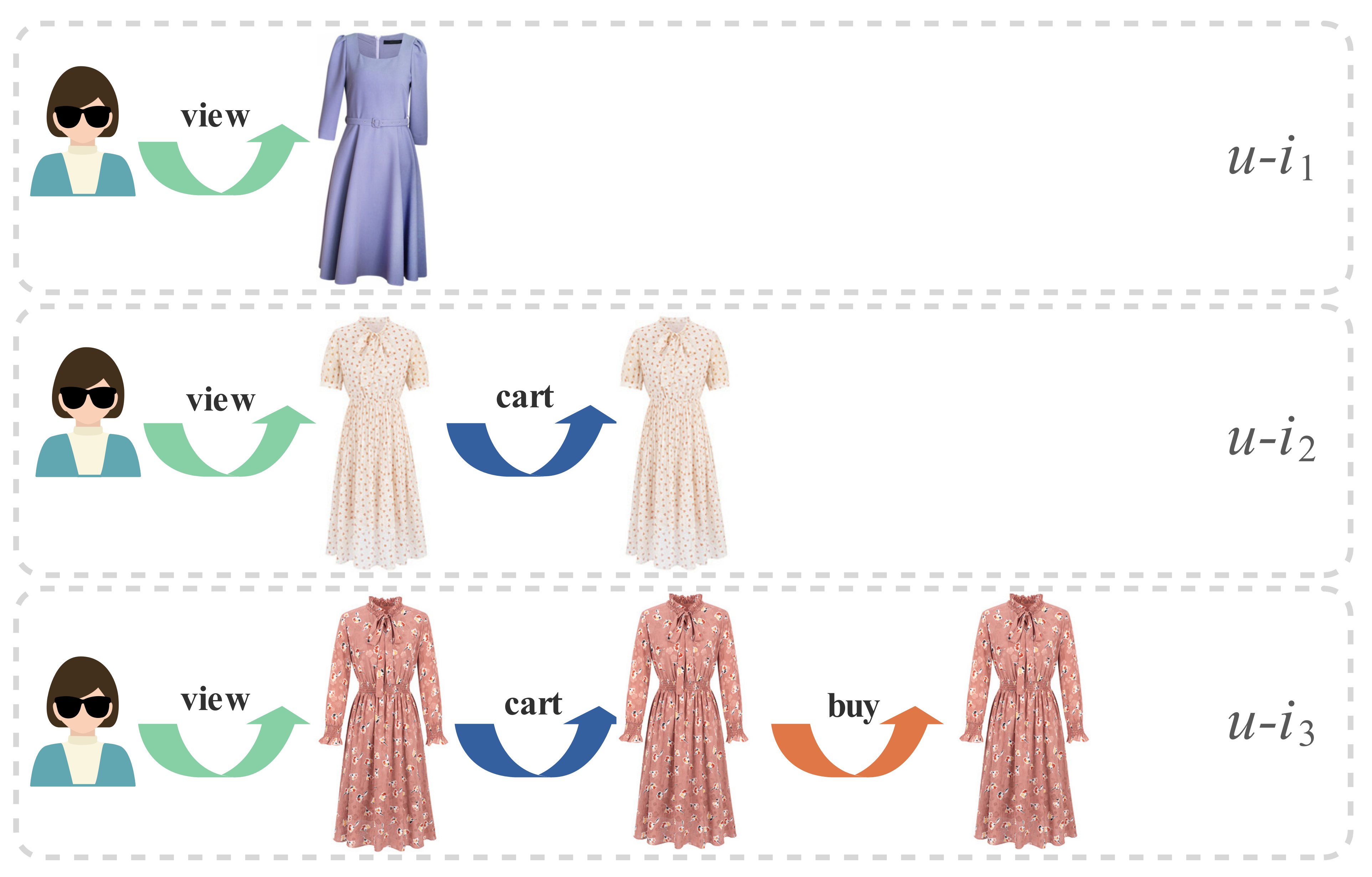}
  \caption{The illustration of the sequence and relation of different types of behaviors.}
  \Description{}
  \label{fig:cascade_show}
\end{figure}

\subsection{CRGCN Model}

In this section, we will introduce our CRGCN model in detail. As discussed in the introduction, we aim to explore the connections between different behaviors to obtain a more complete representation of user and item embeddings. 
In fact, before the target behavior happens between the user and the item, it is often the case that the user has already taken some other behaviors to the item~\cite{ZhouDTY18, WanM18}. 
Taking the e-commerce platform as an example, the possible interactions between users and items under different behaviors are shown in Fig.~\ref{fig:cascade_show}. 
While a user $u$ uses the platform, she may be attracted by the appearance or description of item $i_1$, and thus leads to the \emph{view} behavior. 
However, after viewing the details of the product, she finds out that the item does not match her taste. As a result, she would not take further action on the item, such as \emph{cart}.
Similarly, if she adds $i_2$ to the shopping cart after viewing, the user might not buy it due to price or some other reasons (for example, she finds another item such as $i_3$ can better fit her needs). From this example, we can have the following three observations about the interaction between users and items. First, only partial information of user preferences can be observed from the interaction of a single behavior. Second, along with more behaviors involved in the interaction with an item, more information about user preferences towards the item is revealed. Last, the order of behaviors interacted with the item also indicates a cascading relationship of user preferences among different behaviors.

\begin{figure}[htb]
  \centering
  \includegraphics[width=\textwidth]{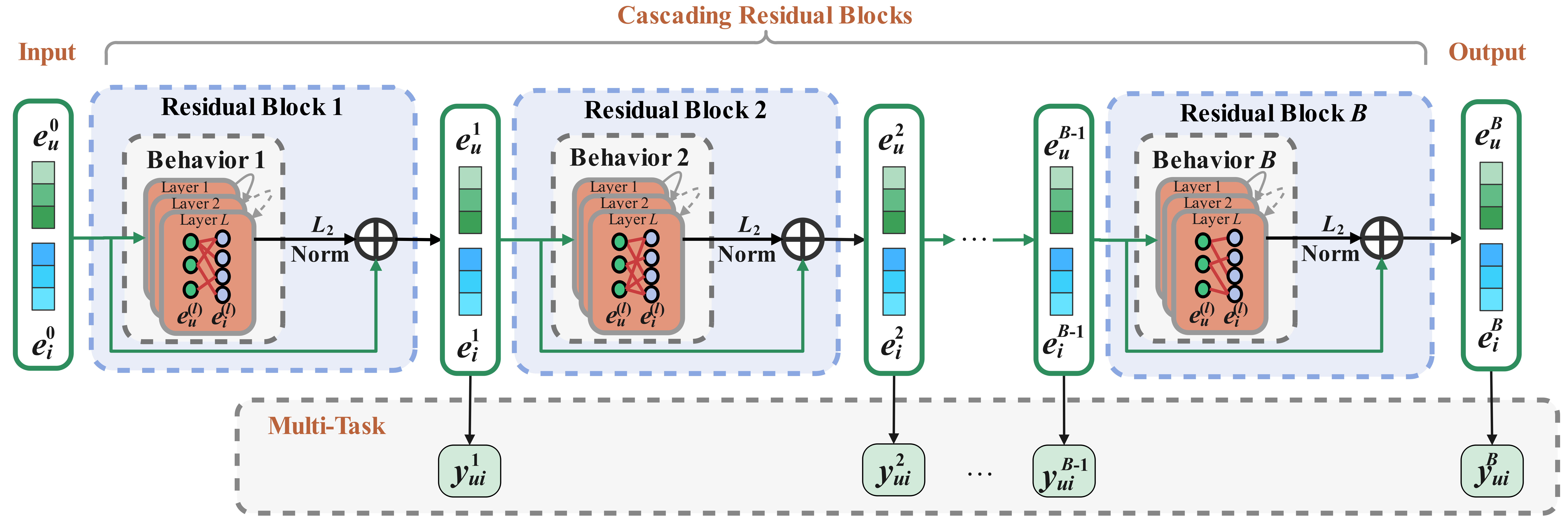}
  \caption{Structure of our CRGCN model.}
  \Description{}
  \label{fig:global}
\end{figure}

Enlighten by the above three points, we attempt to make full use of all behavioral data and design the model to improve the prediction ability of target behavior by well exploiting the relationship among behaviors. The overall structure of our CRGCN model is shown in Fig.~\ref{fig:global}. The key component of our model is the residual block, which not only learns the user and item embeddings from each type of behavior but also captures the cascading effect between different types of behaviors. The influence of the previous behavior on the current behavior can be obtained from the input of the residual module. Therefore, the functionality of our residual block sequence can be divided into two parts: \textit{single behavior modeling} and \textit{cascading effect modeling}. The part of behavior modeling is achieved by each residual block to model the behavioral features from each behavior. The sequence of residual blocks models the cascading effects by leveraging multiple behaviors in a specific order to capture the preference dependency information among different behaviors. Specifically, the captured behavioral features are incrementally integrated into the basic features through a short-cut path, and then fed to the next residual block. It refines the user/item embeddings through information propagation across different behaviors. 
Last, multi-task learning is used to jointly learn the embeddings from different behaviors, improving the ability to predict the target behavior.

\subsubsection{\textbf{Embedding initialization}}
Following the common approach used in existing recommendation methods~\cite{LightGCN, JinG0JL20, GaoHGCFLCYSJ21}, we associate each user and item with an ID embedding. Specifically, let $\boldsymbol{P} \in \mathbb{R}^{M \times d}$ and $\boldsymbol{Q} \in \mathbb{R}^{N \times d}$ be the embedding matrices for the user and item embedding initialization, where  $M$ and $N$ denote the number of users and items, respectively; and $d$ denotes the embedding size. Formally, given the one-hot embedding matrix  $\boldsymbol{ID}^{\mathcal{U}}$ and $\boldsymbol{ID}^{\mathcal{I}}$ for users and items,  the embeddings are initialized as:

\begin{equation}
  \label{eq:init}
  \begin{aligned}
    \boldsymbol{e}_{u}^{0} = \boldsymbol {P} \cdot \boldsymbol{ID}_u^{\mathcal{U}}, \quad
    \boldsymbol{e}_{i}^{0} = \boldsymbol {Q} \cdot \boldsymbol{ID}_i^{\mathcal{I}},
  \end{aligned}
\end{equation}
where $\boldsymbol{e}_{u}^{0}$ and $\boldsymbol{e}_{i}^{0}$ are the initialized user $u$'s and item $i$'s embeddings, respectively. $\boldsymbol{ID}_u^{\mathcal{U}}$ ($\boldsymbol{ID}_i^{\mathcal{I}}$) is user $u$'s (item $i$'s) one-hot vector. It is worth mentioning that the embedding matrices $\boldsymbol{P}$ and $\boldsymbol{Q}$ are the only learnable parameters in our model.

\subsubsection{\textbf{Cascading Residual blocks}}
The goal of cascading residual blocks is to extract user preferences from individual behaviors and also capture the cascading relations of user preferences among behaviors to comprehensively learn user preferences. The main idea is to take the basic features as the initialized user and item embeddings, and then continuously refine them by leveraging the behavioral features learned from each type of behavior. In the next, we will describe the residual block sequence in detail from its two functionalities.

\textbf{Single behavior modeling.}
This part is to learn user preferences from a single behavior (\textit{i.e.}, behavioral features). 
In recent years, Graph Convolutional Networks (GCNs)~\cite{YingHCEHL18, Wang0WFC19, liu2021impgcn,liu2022aagcn, WuYCLH020} have shown a strong ability in learning from graph-structured data, and have demonstrated good recommendation performance. In order to mine the behavioral features contained in the historical interaction data, we employ neighborhood aggregation, which is an essential component in GCN. To be more specific, inspired by the recent advances of GCN-based recommendation~\cite{LightGCN}, we aggregate the information from neighbors without nonlinear transformation.
The information aggregated from neighbors intuitively reflects the user's interests in the current behavior based on learning from high-order connectivity. 
The neighborhood aggregation for updating user node embeddings is formulated as follows:

\begin{equation}
  \label{eq:GCN_original}
  \boldsymbol e_{u}^{(l)} = \textbf{AGG}(\{\boldsymbol e_{i}^{(l-1)}: i \in N_{u}\}), 
\end{equation}
where $\textbf{AGG}$ denotes the aggregation function, which aggregates the information from neighboring nodes of the user $u$; and $\boldsymbol e_{i}^{(l-1)}$ denotes the item embedding of neighboring node of the user $u$ from the (\textit{l}-1)\textit{-th} layer.
Here $e_{u}^{(l)}$ denotes the user embedding in the \textit{l}-th layer, and $N_{u}$ denotes the set of items that are interacted with by the user $u$. 
The embeddings of item nodes are updated in the same way. For simplicity, the aggregation function in standard GCN~\cite{KipfW17} is adopted in our implementation, namely:
\begin{equation}
  \label{eq:agg}
  \begin{aligned}
    \boldsymbol e_{u}^{(l)} &= \sum_{i \in N_{u}} \frac{1}
    {\sqrt{\left\lvert N_{u} \right\rvert} \sqrt{\left\lvert N_{i} \right\rvert}} \boldsymbol e_{i}^{(l-1)}, \\ 
    \boldsymbol e_{i}^{(l)} &= \sum_{u \in N_{i}} \frac{1}
    {\sqrt{\left\lvert N_{i} \right\rvert} \sqrt{\left\lvert N_{u} \right\rvert}} \boldsymbol e_{u}^{(l-1)}, 
  \end{aligned}
\end{equation}
where  $\frac{1}{\sqrt{\left\lvert N_{u} \right\rvert} \sqrt{\left\lvert N_{i} \right\rvert}}$ denotes the normalization coefficient, $\boldsymbol e_{u}^{(l-1)}$ (${\boldsymbol e_{i}^{(l-1)}}$) denotes the user (item) outputs from the (\textit{l}-1)\textit{-th} layer, $e_u^{(l)}$ and $e_i^{(l)}$ are the final output of the GCN (\textit{i.e.}, behavioral features). when $l=1$, $e_u^{0}$ and $e_i^{0}$ denote the input of the GCN, respectively. For the convenience of the description below, we define the final output of GCN as $e'_u$ and $e'_i$ (\textit{i.e.}, $e'_u = e_u^{(l)}, e'_i = e_i^{(l)}$).

In this part, we exploit the neighborhood aggregation of GCN to learn the user preferences (\textit{i.e.}, behavioral features) from each behavior. Remind that in our model, we assume that only a partial preference of users can be observed from a single behavior. In the next, we will introduce how to use these behavioral features to refine user preferences in detail.

\textbf{Cascading effect modeling.}
As discussed above, different behaviors often reflect different aspects of user preferences toward an item. More importantly, the behaviors interacting with items in a certain order reveal user preferences at different degrees. We aim to continuously refine user preferences by integrating all behavioral features and exploiting the connections between different behaviors. We achieve the goal by answering two questions: 1) how to integrate the behavioral features learned from different behaviors? and 2) how to explore the connection between different behaviors?


For the first question, in each block, we can fuse the behavioral features learned in this block with the input (\textit{i.e.}, the output of the previous block) to preserve the features from the previous block. Therefore, we design a residual (\textit{i.e.}, short-cut) connection to refine the input embedding of each behavior. This connection directly connects the feature embeddings learned from the previous block (the input feature of the block) with the one learned from the behavior data in this block. The summation method is then used to merge the two feature embeddings as output. Because the numeric value of features after the GCN learning may be of a different range from the input of the block, the direct summation may cause one feature embedding takes a dominant role in the generated results, making the other one negligible, \textit{i.e.}, when the numeric values of two embeddings are different in the order of magnitude. To avoid this problem, we take a normalization operation on the behavioral features before summation. For simplicity, we adopt $L_2$ normalization in our model: 


\begin{equation}
  \label{eq:norm}
  \begin{aligned}
    \boldsymbol{\tilde{e}}_{u} = \frac{\boldsymbol{e'}_{u}}{\left\lVert \boldsymbol{e'}_{u} \right\rVert_{2}}, \quad
    \boldsymbol{\tilde{e}}_{i} = \frac{\boldsymbol{e'}_{i}}{\left\lVert \boldsymbol{e'}_{i} \right\rVert_{2}}.
  \end{aligned}
\end{equation}

After normalization, we fuse the behavioral features learned in the current block with the one output from the previous block via the residual connection, namely,

\begin{equation}
  \label{eq:fuse}
  \begin{aligned}
    \boldsymbol{e}_{u_{out}} &= \boldsymbol{e}_{u_{in}} + \boldsymbol{\tilde{e}}_{u}, \\
    \boldsymbol{e}_{i_{out}} &= \boldsymbol{e}_{i_{in}} + \boldsymbol{\tilde{e}}_{i},
  \end{aligned}
\end{equation}
where $\boldsymbol{e}_{u_{in}}$ and $\boldsymbol{e}_{i_{in}}$ are the input user and item embeddings of the current block, which are also the output of the previous block. $\boldsymbol{e}_{u_{out}}$ and $\boldsymbol{e}_{i_{out}}$ denote the output user and item embedding of the current block, respectively.

 With the designed residual block, the second problem can be easily addressed by connecting the residual blocks in the order of behaviors, as shown in Fig.~\ref{fig:global}. In this structure, the output of the previous residual block is taken as the input of the next block to deliver the extracted behavioral information from one behavior to the next one. The information delivery between behaviors can bring us two benefits: firstly, from the perspective of learning user preferences, it can continuously refine the embeddings to model user preferences more accurately; secondly, for tackling the data sparsity problem, it can make better use of data that has not been converted into target behaviors to learn user preferences and alleviate the problem of cold-start users to some extent. Concretely,  the embeddings are learned and refined  in the cascading residual blocks as follows:


\begin{equation}
  \label{eq:cascade_result}
  \begin{aligned}
    \boldsymbol{e}^{1} &= \boldsymbol{e}^{0} + \boldsymbol{\tilde{e}}^{1}, \\
    \boldsymbol{e}^{2} &= \boldsymbol{e}^{1} + \boldsymbol{\tilde{e}}^{2}, \\
    &\cdots \\
    \boldsymbol{e}^{B} &= \boldsymbol{e}^{B-1} + \boldsymbol{\tilde{e}}^{B}, \\
  \end{aligned}
\end{equation}
where \textit{B} is the number of the behavior, $\boldsymbol{e}^{B-1}$ denotes the input of the \textit{B-th} residual block (\textit{i.e.}, the output of the (\textit{B}-1)\textit{-th} residual block), $\boldsymbol{\tilde{e}}^{B}$ denotes the normalization behavioral features learned from the \textit{B-th} behavior, and $\boldsymbol{e}^0$ represents the initialized user and item embedding (\textit{i.e.}, basic features). 
Based on this design,  we implement message delivery in different behavior residual blocks from the embedding level. 


Through the above design, the behavioral features are fused into the basic features in an incremental form to refine the user and item embeddings. With the cascading residual blocks, our model explicitly takes the cascading effects between different behaviors into the embedding learning process. Meanwhile, the residual design can ensure that the information can be well preserved and delivered to the next behavior even when the current behavior has no interaction data.
Thus, this helps alleviate the data sparsity and cold start issues. 


\subsubsection{\textbf{Multi-Task learning}} 
Multi-task learning (MTL)~\cite{TangLZG20} is a kind of joint-training paradigm for different-yet-related tasks. In MTL, the performance of each task is improved by updating shared parameters or shared models. 
As for CRGCN, each residual block learns a type of behavioral feature, and all residual blocks share basic features through a cascading structure. 
To ensure effective learning of CRGCN, we take the output of each residual block as a prediction task for the current behavior. 
Thanks to the delivery of information in the cascading structure, during the training of the current task, it can not only train the current residual block but also train the previous ones.

\emph{\textbf{Loss function}}. We design a loss function for each behavior to supervise the learning process of behavioral features. As shown in Fig.~\ref{fig:global}, our CRGCN model can obtain each user' embedding set $\{\boldsymbol{e}_u^1, \boldsymbol{e}_u^2, \cdots, \boldsymbol{e}_u^B\}$ and each item' embedding set $\{\boldsymbol{e}_i^1, \boldsymbol{e}_i^2, \cdots, \boldsymbol{e}_i^B\}$ after learning for each behavior, where $B$ denotes the number of behaviors. 
We then obtain the relevance scores $y_{ui}$ of user-item interaction by calculating the inner product of both as follows:

\begin{equation}
  \label{eq:score}
  y_{ui} = \boldsymbol{e}_{u}^{\top} \boldsymbol{e}_{i}.
\end{equation}

It is necessary to ensure that the score of the observed user-item pair is higher than that of the unobserved one. Given the first type of behavior as an example, the loss function is formulated as follows:

\begin{equation}
  \label{eq:loss_func}
  \mathcal{L}_1 = \sum_{(u,i,j) \in O} -ln \sigma(y_{ui}-y_{uj}),
\end{equation}
where $O=\{(u,i,j)|(u,i) \in \mathcal{R}^{+}, (u,j) \in \mathcal{R}^{-}\}$ is defined as positive and negative sample pairs, and $\mathcal{R}^{+}$ ($\mathcal{R}^{-}$) denotes the sample that has been observed (unobserved) in the current behavior. Here $\sigma(\cdot)$ denotes the sigmoid function. 
The loss function for other behaviors is similar. 
We then can get the set of loss functions $\mathcal{B} = \{\mathcal{L}_{1}, \cdots, \mathcal{L}_{b}, \cdots, \mathcal{L}_{B}\}$, where $b$ is the \textit{b-th} behavior. 
Based on MTL, we treat the learning of each behavior as a task. 
The final loss is an aggregation of all the losses across different behaviors. It is formulated as follows:

\begin{equation}
  \label{eq:total_loss}
  \mathcal{L} = \sum_{i \in \mathcal{B}} \mathcal{L}_i + \beta \cdot \left\lVert \boldsymbol \Theta \right\rVert_{2},
\end{equation} 
where $\boldsymbol \Theta$ represents all trainable parameters in our model and $\beta$ is the coefficient that controls the strength of the $L_2$ normalization to prevent over-fitting. 

The direct optimization of the loss function will learn the parameters of multiple tasks.
Specifically, it updates the initialized embeddings (\textit{i.e.}, basic features) from the perspective of multiple tasks, directly and indirectly, since initialized embeddings are the main learnable parameters in our model. 

\emph{\textbf{Training}}. We implement our model on Pytorch\footnote{https://pytorch.org/} and adopt Adam~\cite{Adam} for optimization. The mini-batch training strategy is also used to speed up the training process. To generate a mini-batch, we sample the interaction data of different behaviors on a user-by-user basis to ensure that each user is trained. Specifically, given a user in a batch, a positive-negative pair is sampled for each behavior. The sampling is in the form of a triple $(u, i, j)$, where $u$ represents the user, $i$ is the positive sample, and $j$ is the negative one. Take one type of behavior as an example, where $i$ is a data sample randomly from the observed interaction data. When user $u$ has no interaction data under this behavior, a default triple (\textit{i.e.}, $(0, 0, 0)$) is returned  to keep the training running properly.

In order to avoid the over-fitting problem,  two widely-used dropout strategies are adopted in experiments~\cite{BergKW17, Wang0WFC19, JinG0JL20}: message dropout and node dropout. 
Specifically, message dropout is used to drop out the information in the embedding, and node dropout is used to randomly drop out nodes in the graph.

\subsubsection{\textbf{Complexity analysis}} \label{Complexity}
Our CRGCN model does not introduce any trainable parameters other than the ones in user and item embedding initialization. Thus, CRGCN  has the same trainable parameters as the basic MF~\cite{RendleFGS09} model, demonstrating a big advantage in space complexity. The computing complexity of our model is analyzed in the following. 
The time cost of our model is mainly from computing the adjacency matrix, graph convolution, and BPR loss. Let $|E|$ be the number of edges in the user-item interaction graph, $B$ denotes the number of behavior types, $n$ is the number of epochs, $b$ denotes the size of each training batch, $d$ represents the embedding size, and $L$ represents the number of GCN layers. In the process of learning the adjacency matrix, the computational complexity is $\mathcal{O}(2|E|B)$. In the Graph convolution process, the computational complexity is $\mathcal{O}(2|E|nBLd\frac{|E|}{b})$. The computational complexity for BPR Loss is $\mathcal{O}(2|E|nBd)$. Due to the number of GCN layers in our model is 1 (i.e., $L=1$), the total computing complexity of CRGCN is $\mathcal{O}(2|E|B + 2|E|nBd\frac{|E|}{b} + 2|E|nBd)$. Since the number of behaviors in multi-behavior tasks is usually very small ($B=4$ in Tmall and Jdata dataset, and $B=3$ in Beibei dataset). Thus, the computing complexity of CRGCN is similar to LightGCN, which is $\mathcal{O}(2|E| + 2|E|nLd\frac{|E|}{b} + 2|E|nd)$. Compared with other multi-behavior recommendation models, our model enjoys a big advantage in the computation complexity.

\subsection{Model Discussion}

In this section, we will analyze and discuss the design of each part in our model to justify the rationality of the design of CRGCN. The overall design of our model is to explore the relationship between different behaviors, and achieve the continuous refinement of user preferences with the exploitation of different behaviors. 

\begin{itemize}
  \item \textbf{Behavior modeling}. The design of this part aims to mine the user preferences from different types of behaviors. In recent years, GCN has set up a new standard for collaborative filtering-based recommendation methods due to its powerful representation learning ability from the graph structure. In the recommendation scenario, the GCN methods can leverage the high-order connectivity in the user-item bipartite graph for the user and item embedding learning. Accordingly, we also adopt the GCN techniques in our model to learn user and item embeddings from the interaction data of each behavior.
  
  \item \textbf{Residual block}. Residual blocks are designed to integrate user preferences learned from different behaviors. A short-cut connection is used to preserve the behavioral features learned from the previous block. Before the fusion, $L_2$ normalization is performed on the behavioral features learned by GCN to balance the effects of the behavioral features learned from two adjacent blocks. 
  
  \item \textbf{Cascading residual blocks}. The cascading structure connects all residual blocks in a certain order, and takes the output of the former as the input of the latter. Such a design connects all behaviors to deliver behavioral features for embedding refinement. This design enables our model to exploit the connections between different behaviors for embedding learning explicitly. 
  
  \item \textbf{Multi-Task learning}. We adopt the multi-task learning strategy by treating the learning of each behavioral feature as an individual task. From the local view of a single task, it can well utilize the current behavior information to learn user preferences; and from the global view, different tasks are interacted with each other and learned jointly together with the cascading structure, enhancing user preferences learning and refinement across different types of behaviors.
\end{itemize}

\section{experiment} \label{experiment}

To evaluate the effectiveness of our CRGCN model, we conduct comprehensive experiments on three publicly available datasets, which are commonly used to evaluate the multi-behavior recommendation models. In particular, we aim to answer the following research questions:
\begin{itemize}
  \item \textbf{RQ1:} How does our CRGCN model perform as compared with the state-of-the-art recommendation models that are learned from single- and multi-behavior data?
  \item \textbf{RQ2:} How does each module in our CRGCN model affect the recommendation performance?
   \item \textbf{RQ3:} How does the multi-behavior information (e.g., the number or the order of the behaviors) impact the recommendation performance?
  \item \textbf{RQ4:} Can our model effectively leverage the multi-behavior information to alleviate the cold-start users problem as compared with the existing multi-behavior recommendation models?
  \item \textbf{RQ5:} How about the computing efficiency of our CRGCN model?
\end{itemize}

\subsection{Experiment Settings}

\subsubsection{\textbf{Dataset}}

Three public real-world datasets have been adopted for experiments: Tmall\footnote{https://tianchi.aliyun.com/dataset/dataDetail?dataId=649}, Beibei\footnote{https://www.beibei.com}, and Jdata\footnote{https://jdata.jd.com/html/detail.html?id=8}.

\begin{itemize}
  \item \textbf{Tmall.} This dataset is collected from Tmall\footnote{https://www.tmall.com/}, one of the largest e-commerce platforms in China. It contains 41,738 users and 11,953 items with 4 types of behaviors, \textit{i.e.}, \emph{view}, \emph{collect}, \emph{cart}, and \emph{buy}. On the Tmall platform, users can buy the item directly after viewing, or add it to the cart before purchasing, or they may just click on the collection instead of the \textit{buy} behavior. 

  \item \textbf{Beibei.} This dataset is collected from Beibei\footnote{https://www.beibei.com/}, the largest infant product retail e-commerce platform in China. This dataset contains 21,716 users and 7,977 items with three types of behaviors, including \emph{view}, \emph{cart}, and \emph{buy} behavior data within the period from 2017/06/01 to 2017/06/30. On the Beibei platform, users' shopping process is carried out according to the process of \emph{view}, \emph{cart}, and finally \emph{buy}.
  
  \item \textbf{Jdata.} This dataset is collected from JD\footnote{https://www.jd.com/}, a comprehensive online retailer in China and one of the most popular and influential e-commerce websites in the Chinese e-commerce field. This dataset contains 93,334 users and 24,624 items with 4 types of behaviors, \textit{i.e.}, \emph{view}, \emph{collect}, \emph{cart}, and \emph{buy} behavior data within the period from 2018/02/01 to 2018/04/15. The behavior is similar to that of Tmall.

\end{itemize} 
For the three datasets, we followed the previous work to merge the duplicated user-item interactions by keeping the earliest one~\cite{GaoHGCFLCYSJ21, JinG0JL20}. The statistical information of the three datasets used in our experiments is summarized in Table~\ref{tab:dataset}.

\begin{table}[htb]
  \caption{Statistics of three real-world benchmark datasets.}
  \label{tab:dataset}
    \begin{tabular}{ccccccc}
      \toprule
      \textbf{Dataset} & \textbf{Users} & \textbf{Items} & \textbf{Buy} & \textbf{Cart} & \textbf{Collect} & \textbf{View} \\
      \midrule
      \textbf{Tmall}  & 41,738 & 11,953 & 255,586 & 1,996   & 221,514 & 1,813,498 \\
      \textbf{Beibei} & 21,716 & 7,997  & 304,576 & 642,622 & -       & 2,412,586 \\
      \textbf{Jdata}  & 93,334 & 24,624 & 333,383 & 49,891  & 45,613  & 1,681,430 \\
      \bottomrule
    \end{tabular}
\end{table}
\subsubsection{\textbf{Evaluation Protocols}}
We adopt the widely used leave-one-out strategy for evaluation~\cite{LightGCN, GaoHGCFLCYSJ21, HeLZNHC17}, which means for each user, the test set is comprised of one positive item and all the items that she has not interacted with before. In the training stage, the last positive item for each user is selected to construct the validation set for hyper-parameter tuning. In the evaluation stage, all the items in the test set are ranked according to the predicted scores by recommendation algorithms. We sort the items by predicting user preferences for all the items that do not appear in the training set, and the top-$K$ ranked items will be used for evaluation. Two popular evaluation metrics in recommendation  HR@K and NDCG@K  are adopted to evaluate the performance:
\begin{itemize}
  \item \textbf{HR@K:} Hit Ratio (HR) is a commonly used metric to measure whether the positive test item is recommended in the top $K$ items in the ranking list.
  \item \textbf{NDCG@K:} Normalized Discounted Cumulative Gain (NDCG) takes the position of correctly recommended items into consideration by assigning a higher score to the hit at a higher position. 
\end{itemize}

\subsubsection{\textbf{Baselines}}
We compare our CRGCN model with several competitive recommendation models, including two single-behavior methods and six multi-behavior models. We briefly introduce these methods as follows.

\textbf{Single-behavior model:}

\begin{itemize}
  \item \textbf{MF-BPR}~\cite{RendleFGS09}. MF-BPR makes recommendations based on a single behavior and has been widely used as a baseline to examine the performance of newly proposed models. BPR is a widely used optimization strategy, which assumes that the predicted scores of positive samples are higher than those of negative samples.  

  \item \textbf{LightGCN}~\cite{LightGCN}. It becomes a new standard for CF models by exploiting a single user-item interaction behavior in recommendation. LightGCN leverages the GCN technique to exploit the high-order connectives in the user-item bipartite graph for recommendation. In particular, it removes the feature transformation and non-linear activation components in traditional GCN models to simplify the model structure and achieves a significant performance improvement over its counterpart.
\end{itemize}

\textbf{Multi-behavior model:}

\begin{itemize}
  \item \textbf{R-GCN}~\cite{SchlichtkrullKB18}. R-GCN differentiates the relations between nodes via edge types in the graph and designs different propagation layers for different types of edges to model the relation information. This model can adapt to the task of multi-behavior recommendation.

  \item \textbf{NMTR-NCF}~\cite{GaoHGCFLCYSJ21}. It is a deep learning model for multi-behavior recommendation. NMTR-NCF develops a neural network model to capture the complicated and multi-type interactions between users and items. It sequentially passes the interaction score of the current behavior to the next and also adopts multi-task learning to jointly optimize shared parameters.
  
  \item \textbf{NMTR-GCN}. This is a modified NMTR model which uses single-layer GCN to replace the Neural Collaborative Filtering (NCF) module~\cite{HeLZNHC17} in NMTR-NCF. 

  \item \textbf{MBGCN}~\cite{JinG0JL20}. This model constructs a unified multi-behavior graph to learn user preferences through the user-item propagation layer and employs learnable parameters to assign weights for different behaviors during layer aggregation. In addition, it also exploits the high-order item-item relations to enhance the item embedding learning. 

  \item \textbf{GNMR}~\cite{XiaHXDLB21}. This model designs a relation aggregation network to model interaction heterogeneity. It attempts to explore the dependencies among different types of behaviors via recursive embedding propagation over the multi-behavior interaction graph.

  \item \textbf{S-MBRec}~\cite{GuWSX22}. This model consists of supervised and self-supervised learning tasks. It uses multiple GCNs to learn the user and item embeddings from each behavior and adopts a star-style contrastive learning strategy, which constructs a contrastive view pair for the target and each auxiliary behavior.

\end{itemize}

\subsubsection{\textbf{Hyper-parameter Settings}} \label{Settings}
We implemented our CRGCN model in Pytorch\footnote{https://pytorch.org/}. The source code of our implementation is released~\footnote{The source codes are available at https://github.com/MingshiYan/CRGCN.}. In our experiments, the mini-batch size of all models is set to 1024, and the embedding size is fixed to 64~\cite{LightGCN}. For all the models using pair-wise learning loss~\cite{GaoHGCFLCYSJ21}, we randomly sampled 4 negative samples for each positive sample~\cite{HeLZNHC17, GaoHGCFLCYSJ21}. In addition, we used the grid search to tune the learning rate in the range of $[1e^{-2}, 3e^{-3}, 1e^{-3}, 1e^{-4}]$ and tuned the regularization weight (\textit{i.e.}, $\beta$) in the range of $[1e^{-2}, 1e^{-3}, 3e^{-4}, 1e^{-4}]$. For the other hyper-parameters in the baselines, we carefully tuned them according to their original papers. Furthermore, we adopted an early stop strategy in the training stage, that is, the training process will be stopped when HR@20 on the validation set does not increase within 20 epochs. Note that the NMTR-NCF~\cite{GaoHGCFLCYSJ21} model needs the behaviors to be happening in a certain order; we strictly followed the strategy reported in the original paper on the Tmall dataset, \textit{i.e.} only \emph{view} and \emph{buy} behavior are used.

\begin{table*}[htb]
  \caption{Overall performance comparisons on Tmall dataset.}
  \label{tab:tmall}
  \resizebox{\textwidth}{!}{
  \begin{tabular}{cccccccccc}
  \toprule
  & \textbf{Method} & \textbf{HR@10} & \textbf{NDCG@10} & \textbf{HR@20} & \textbf{NDCG@20} & \textbf{HR@50} & \textbf{NDCG@50} & \textbf{HR@80} & \textbf{NDCG@80} \\
  \hline
  \multirow{2}{*}{\textbf{One-behavior}} & MF-BPR & 0.0230 & 0.0124 & 0.0316 & 0.0144 & 0.0434 & 0.0166 & 0.0541 & 0.0183 \\
  & LightGCN & 0.0393 & 0.0209 & 0.0538 & 0.0243 & 0.0813 & 0.0295 & 0.0984 & 0.0322 \\
  \hline
  \multirow{6}{*}{\textbf{Multi-behavior}} & R-GCN & 0.0316 & 0.0157 & 0.0489 & 0.0198 & 0.0826 & 0.0262 & 0.1067 & 0.0300   \\
  & NMTR-NCF & 0.0517 & 0.0250  & 0.0847 & 0.0330  & 0.1498 & 0.0456 & \underline{0.1963} & 0.0531 \\
  & NMTR-GCN & 0.0536 & 0.0286 & 0.0721 & 0.0330  & 0.1037 & 0.0391 & 0.1256 & 0.0426 \\
  & MBGCN & 0.0549 & 0.0285 & 0.0799 & 0.0345 & 0.1285 & 0.0438 & 0.1629 & 0.0493 \\
  & GNMR & 0.0393 & 0.0193 & 0.0619 & 0.0247 & 0.1071 & 0.0332 & 0.1410 & 0.0388 \\
  & S-MBRec & \underline{0.0694} & \underline{0.0362} & \underline{0.1009} & \underline{0.0438} & \underline{0.1553} & \underline{0.0544} & 0.1901 & \underline{0.0601} \\
  & \textbf{CRGCN} & \textbf{0.0840} & \textbf{0.0442} & \textbf{0.1238} & \textbf{0.0540} & \textbf{0.1994} & \textbf{0.0685} & \textbf{0.2491} 
  & \textbf{0.0766}    \\
  \hline
  & Improvement & 21.04\% & 22.10\% & 22.70\% & 23.29\% & 28.34\% & 25.92\% & 26.90\% & 27.45\% \\
  \bottomrule
  \end{tabular}
  }
\end{table*}

\begin{table*}[htb]
  \caption{Overall performance comparisons on Beibei dataset.}
  \label{tab:beibei}
  \resizebox{\textwidth}{!}{
  \begin{tabular}{cccccccccc}
  \toprule
   & \textbf{Method} & \textbf{HR@10} & \textbf{NDCG@10} & \textbf{HR@20} & \textbf{NDCG@20} & \textbf{HR@50} & \textbf{NDCG@50} & \textbf{HR@80} & \textbf{NDCG@80}  \\
  \hline
  \multirow{2}{*}{\textbf{One-behavior}} & MF-BPR & 0.0268  & 0.0139 & 0.0427 & 0.0179 & 0.0793 & 0.0250 & 0.1075 & 0.0297  \\
  & LightGCN & 0.0309 & 0.0161 & 0.0478 & 0.0204 & 0.0880 & 0.0282 & 0.1220 & 0.0339 \\
  \hline
  \multirow{6}{*}{\textbf{Multi-behavior}} & R-GCN & 0.0327 & 0.0161 & 0.0561 & 0.0219  & 0.1118 & 0.0329 & 0.1603 & 0.0409 \\
  & NMTR-NCF & 0.0315 & 0.0146 & 0.0587 & 0.0214 & 0.1276 & 0.0348 & \underline{0.1877} & 0.0448 \\
  & NMTR-GCN & 0.0301 & 0.0144 & 0.0524 & 0.0200 & 0.1139 & 0.0322 & 0.1607 & 0.0399 \\
  & MBGCN & 0.0373 & 0.0193 & 0.0639 & 0.0259 & \underline{0.1287} & 0.0386 & 0.1807 & 0.0472 \\
  & GNMR & 0.0396 & 0.0219 & 0.0640 & 0.0280 & 0.1219 & 0.0394 & 0.1739 & \underline{0.0480} \\
  & S-MBRec & \underline{0.0489} & \underline{0.0253} & \underline{0.0770} & \underline{0.0324} & 0.1234 & \underline{0.0415} & 0.1570 & 0.0471 \\
  & \textbf{CRGCN} & \textbf{0.0539} & \textbf{0.0259} & \textbf{0.0944} & \textbf{0.0361} & \textbf{0.1817} & \textbf{0.0532} & \textbf{0.2536} & \textbf{0.0652} \\
  \hline 
  & Improvement & 10.22\% & 2.37\% & 22.60\% & 11.42\% & 41.18\% & 28.19\% & 35.11\% & 35.83\% \\
  \bottomrule
  \end{tabular}
  }
\end{table*}

\begin{table*}[htb]
  \caption{Overall performance comparisons on Jdata dataset.}
  \label{tab:jdata}
  \resizebox{\textwidth}{!}{
  \begin{tabular}{cccccccccc}
  \toprule
   & \textbf{Method} & \textbf{HR@10} & \textbf{NDCG@10} & \textbf{HR@20} & \textbf{NDCG@20} & \textbf{HR@50} & \textbf{NDCG@50} & \textbf{HR@80} & \textbf{NDCG@80}  \\
  \hline
  \multirow{2}{*}{\textbf{One-behavior}} & MF-BPR & 0.1850 & 0.1238 & 0.2192 & 0.1325 & 0.2652 & 0.1417 & 0.2890 & 0.1456 \\
  & LightGCN & 0.2252 & 0.1436 & 0.2825 & 0.1582 & 0.3658 & 0.1747 & 0.4108 & 0.1822 \\
  \hline
  \multirow{6}{*}{\textbf{Multi-behavior}} & R-GCN & 0.2406 & 0.1444 & 0.3418 & 0.1588 & 0.4873 & 0.1891 & 0.5548 & 0.2008 \\
  & NMTR-NCF & 0.3142 & 0.1717 & 0.4086 & 0.1966 & 0.5227 & 0.2198 & 0.5843 & 0.2304 \\
  & NMTR-GCN & 0.3190 & 0.1914 & 0.4071 & 0.2006 & 0.5375 & 0.2274 & 0.5926 & 0.2469 \\
  & MBGCN & 0.2803 & 0.1572 & 0.3603 & 0.1790 & 0.5045 & 0.1984 & 0.5741 & 0.2098 \\
  & GNMR & 0.3068 & 0.1581 & 0.3694 & 0.1944 & 0.4607 & 0.2029 & 0.5106 & 0.2114 \\
  & S-MBRec & \underline{0.4125} & \underline{0.2779} & \underline{0.4957} & \underline{0.2989} & \underline{0.6036} & \underline{0.3203} & \underline{0.6584} & \underline{0.3295} \\
  & \textbf{CRGCN} & \textbf{0.5001} & \textbf{0.2914} & \textbf{0.6190} & \textbf{0.3225} & \textbf{0.7685} & \textbf{0.3535} & \textbf{0.8359} & \textbf{0.3652} \\
  \hline 
  & Improvement & 21.24\% & 4.86\% & 24.87\% & 7.90\% & 27.32\% & 10.37\% & 26.96\% & 10.83\% \\
  \bottomrule
  \end{tabular}
  }
\end{table*}

\subsection{Overall Performance (RQ1)}

In this section, we report the performance comparisons between our CRGCN model and the baselines. The experimental results on the three datasets are shown in Table~\ref{tab:tmall}, Table~\ref{tab:beibei}, and Table~\ref{tab:jdata}. Overall, CRGCN achieves the best performance. It can be seen that the CRGCN model significantly outperforms all baselines on three datasets. For the two metrics, the average improvement across different ranges of top K (K$=\{10,20,50,80\}$) items over the second best method can achieve $24.76\%$ and $24.69\%$ on Tmall, $27.28\%$ and $19.45\%$ on Beibei, and $25.10\%$ and $8.49\%$ on Jdata for HR@K and NDCG@K metrics, respectively. This is a remarkable improvement in the recommendation accuracy, demonstrating the effectiveness of our CRGCN model. The contribution of different components in our model will be further analyzed in the ablation study.

For the methods that only leverage a single type of behavior, LightGCN achieves consistently better performance over MF-BPR, demonstrating the advantages of GCN models in taking advantage of the higher-order neighbors' information during the user and item embedding learning. The graph is natural for modeling complex relations and is thus convenient for modeling multi-behavior interactions between users and items. Meanwhile, GCN is powerful in learning node features in graph structures, facilitating the advancement of GCN-based multi-behavior recommendation models. Our model also adopts this approach. Besides R-GCN, all the multi-behavior recommendation models perform better than the single-behavior methods, which demonstrates the potential of exploiting multi-behavior information in user preference modeling. Different from other multi-behavior recommendation models, R-GCN is not designed for this task. R-GCN separately models each behavior and then fuses the features from different behaviors without distinguishing their contributions to the target behavior. Thus it cannot well exploit different behaviors for user preference modeling, resulting in relatively worse performance. MBGCN considers the contributions of different behaviors before fusion and achieves better performance than R-GCN, indicating the importance of considering the contributions of different behaviors in modeling. GNMR attempts to apply the attention mechanism to capture the dependencies between different behaviors. However, it ignores the impact of the order of behaviors, which conveys important information about user preferences on an item. It is interesting to observe that GNMR performs the third best in Beibei, but it does not perform well on the Tmall and Jdata datasets. This is because of the strict requirement on the order of behaviors on Beibei. This requirement reinforces the dependency effects between different behaviors, and is also beneficial for the model to capture the dependency relations. In contrast, the order of behaviors in Tmall is somewhat arbitrary, which limits the performance of GNMR. S-MBRec performs the second best over the three datasets, attributing to the contrastive learning to mine the commonalities between auxiliary behavior and target behavior.


The NMTR model considers the cascading effects of behaviors (\textit{e.g.}, \emph{view}->\emph{cart}->...). It models the effects by passing the prediction scores of a previous behavior to the next one. The NMTR-NCF is the original model proposed in ~\cite{GaoHGCFLCYSJ21}. As we can see, it can achieve the better performance among all the baselines on Tmall and Jdata datasets because of the consideration of the cascading effects of user behaviors. Our CRGCN also considers the cascading effects in the modeling. The core difference between CRGCN and NMTR is in the way of modeling the cascading effects: NMTR models the effects by passing the prediction score while CRGCN captures the effects by passing the embedding learned from the previous behavior to the next one for further refining, which is similar to an embedding refinement process according to users' behavior sequences. Another difference is that CRGCN models the individual behaviors via GCN while NMTR models each behavior by using NCF as the backbone model. To analyze whether the improvement of CRGCN over NMTR is attributed to the use of GCN in modeling each behavior, we replace the NCF with GCN in NMTR (denoted as NMTR-GCN as shown in the tables). It can be seen that NMTR-GCN  can only perform comparable to NMTR-NCF or even slightly worse than NMTR-NCF in most cases. This demonstrates that simply using GCN to replace the NCF cannot achieve performance improvement. The substantial improvement of our CRGCN model should be credited to the way of modeling the cascading effects. The contributions of some specific components in the modeling (\textit{e.g.}, residual block, $L_2$ normalization, one-layer GCN, etc.) will be further analyzed in the next section.

Overall, we can have the following conclusions based on the performance comparisons among all the adopted methods: 1) multi-behavior information is very useful for preference modeling and can help recommendation models make much more accurate predictions; 2) the cascading effect of different behaviors is important for multi-behavior recommendation methods to accurately model user preferences; 3) the way of modeling the cascading effects is also important and can make a big difference on the recommendation performance.  



\subsection{Ablation Study (RQ2)}

To further study the CRGCN model, we conduct extensive ablation studies to examine the effectiveness of different components for the final performance. We analyze the contributions of each component from the following aspects.

\subsubsection{\textbf{Effects of the Residual Block}}


\begin{table}[htb] 
  \caption{Effects of different designs in our residual block ("$L_2$", "SC" represent $L_2$ normalization for behavioral feature and short-cut connection, respectively).}
  \label{tab:ablation_residual}
  \resizebox{\textwidth}{!}{
  \begin{tabular}{ccccccccccccccc}
  \toprule
  \multicolumn{2}{c}{\textbf{Ablation}} & \multicolumn{4}{c}{\textbf{Tmall}} & \multicolumn{4}{c}{\textbf{Beibei}} & \multicolumn{4}{c}{\textbf{Jdata}} \\ \cmidrule(l){1-2} \cmidrule(l){3-6} \cmidrule(l){7-10} \cmidrule(l){11-14}
  \textbf{\textit{L}}$_2$ & \textbf{SC} & \textbf{HR@10} & \textbf{NDCG@10} & \textbf{HR@20} & \textbf{NDCG@20} & \textbf{HR@10} & \textbf{NDCG@10} & \textbf{HR@20} & \textbf{NDCG@20} & \textbf{HR@10} & \textbf{NDCG@10} & \textbf{HR@20} & \textbf{NDCG@20} \\
  \hline
  \Checkmark &            & 0.0137 & 0.0065 & 0.0225 & 0.0086 & 0.0130 & 0.0065 & 0.0249 & 0.0095 & 0.1566 & 0.0868 & 0.2046 & 0.0999 \\
             & \Checkmark & 0.0729 & 0.0375 & 0.1059 & 0.0455 & 0.0350 & 0.0178 & 0.0615 & 0.0244 & 0.3965 & 0.2320 & 0.5093 & 0.2613 \\
  \Checkmark & \Checkmark & \textbf{0.0840} & \textbf{0.0442} & \textbf{0.1238} & \textbf{0.0540} & \textbf{0.0539} & \textbf{0.0259} & \textbf{0.0944} & \textbf{0.0361} & \textbf{0.5001} & \textbf{0.2914} & \textbf{0.6190} & \textbf{0.3225} \\
  \bottomrule
  \end{tabular}
  }
\end{table}


\begin{table}[htb] 
  \caption{Effects of $L_2$ normalization before feature fusion ("-\textit{plain}", "-\textit{LW}", "-$L_2$" represent without $L_2$ normalization, assign learnable weights for the two types of feature, and $L_2$ normalization for behavioral feature, respectively).}
  \label{tab:fusion_compare}
  \resizebox{\textwidth}{!}{
  \begin{tabular}{cccccccccccccc}
  \toprule
  \multirow{2}{*}{\textbf{Model}} & \multicolumn{4}{c}{\textbf{Tmall}} & \multicolumn{4}{c}{\textbf{Beibei}} & \multicolumn{4}{c}{\textbf{Jdata}} \\ \cmidrule(l){2-5} \cmidrule(l){6-9} \cmidrule(l){10-13}
                                  & \textbf{HR@10} & \textbf{NDCG@10} & \textbf{HR@20} & \textbf{NDCG@20} & \textbf{HR@10} & \textbf{NDCG@10} & \textbf{HR@20} & \textbf{NDCG@20} & \textbf{HR@10} & \textbf{NDCG@10} & \textbf{HR@20} & \textbf{NDCG@20} \\
  \hline
  CRGCN-\textit{plain} & 0.0729 & 0.0375 & 0.1059 & 0.0455 & 0.0350 & 0.0178 & 0.0615 & 0.0244 & 0.3965 & 0.2320 & 0.5093 & 0.2613 \\
  CRGCN-\textit{LW}    & 0.0735 & 0.0373 & 0.1068 & 0.0454 & 0.0370 & 0.0191 & 0.0643 & 0.0260 & 0.4419 & 0.2548 & 0.5597 & 0.2855 \\
  \textbf{CRGCN-$L_2$} & \textbf{0.0840} & \textbf{0.0442} & \textbf{0.1238} & \textbf{0.0540} & \textbf{0.0539} & \textbf{0.0259} & \textbf{0.0944} & \textbf{0.0361} & \textbf{0.5001} & \textbf{0.2914} & \textbf{0.6190} & \textbf{0.3225} \\ 
  \bottomrule
  \end{tabular}
  }
\end{table}

The residual block is a core design in our model to learn user preferences by exploiting the multi-behavior in sequence. In particular, user preferences extracted from a behavior are fed into the GCN module of the next behavior to learn more accurate user preferences. In this process, we introduce two special designs: 1) $L_2$ normalization to balance the feature embeddings learned from previous behaviors and the one learned from current behaviors, and 2) a short-cut connection to preserve the features learned from previous behaviors. To analyze the effectiveness of both designs, we carry out experiments on the following two variants of our model as competitors: 1) removing the short-cut connection in the residual block; 2) removing the $L_2$ normalization from CRGCN.

The experimental results on Tmall, Beibei, and Jdata datasets are reported in Table~\ref{tab:ablation_residual}. From the results, we can see that the performance drops sharply after removing the short-cut connection (the first row in Table~\ref{tab:ablation_residual}). This demonstrates the importance of the residual design, which can well preserve the features extracted from previous behaviors and integrate them into the next behaviors to refine user preferences. After removing the $L_2$ normalization, the performance has also greatly decreased. This is because, without normalization, the value of the features learned from GCN might have deviated far from the feature values passed from the residual block of the previous behavior (\textit{e.g.}, in different orders of magnitude). As a result, this will largely weaken the effects of the features passed from the short-cut connection in the network. To further validate the effects of $L_2$ normalization, we conduct another experiment to replace it with learnable weights (denoted as CRGCN-$LW$) for feature fusion. The experimental results are shown in Table~\ref{tab:fusion_compare}. Compared with the direct fusion (\textit{i.e.}, CRGCN-$plain$), the use of learnable weights can yield slight improvement, while there is still an enormous gap compared to our model with $L_2$ normalization. This validates the importance of ensuring the feature values are in the same numerical range before feature fusion. In a nutshell, the above experimental results can well validate the effectiveness of our designs in the residual block.

\begin{figure}[htb]
  \centering
  \includegraphics[width=\textwidth]{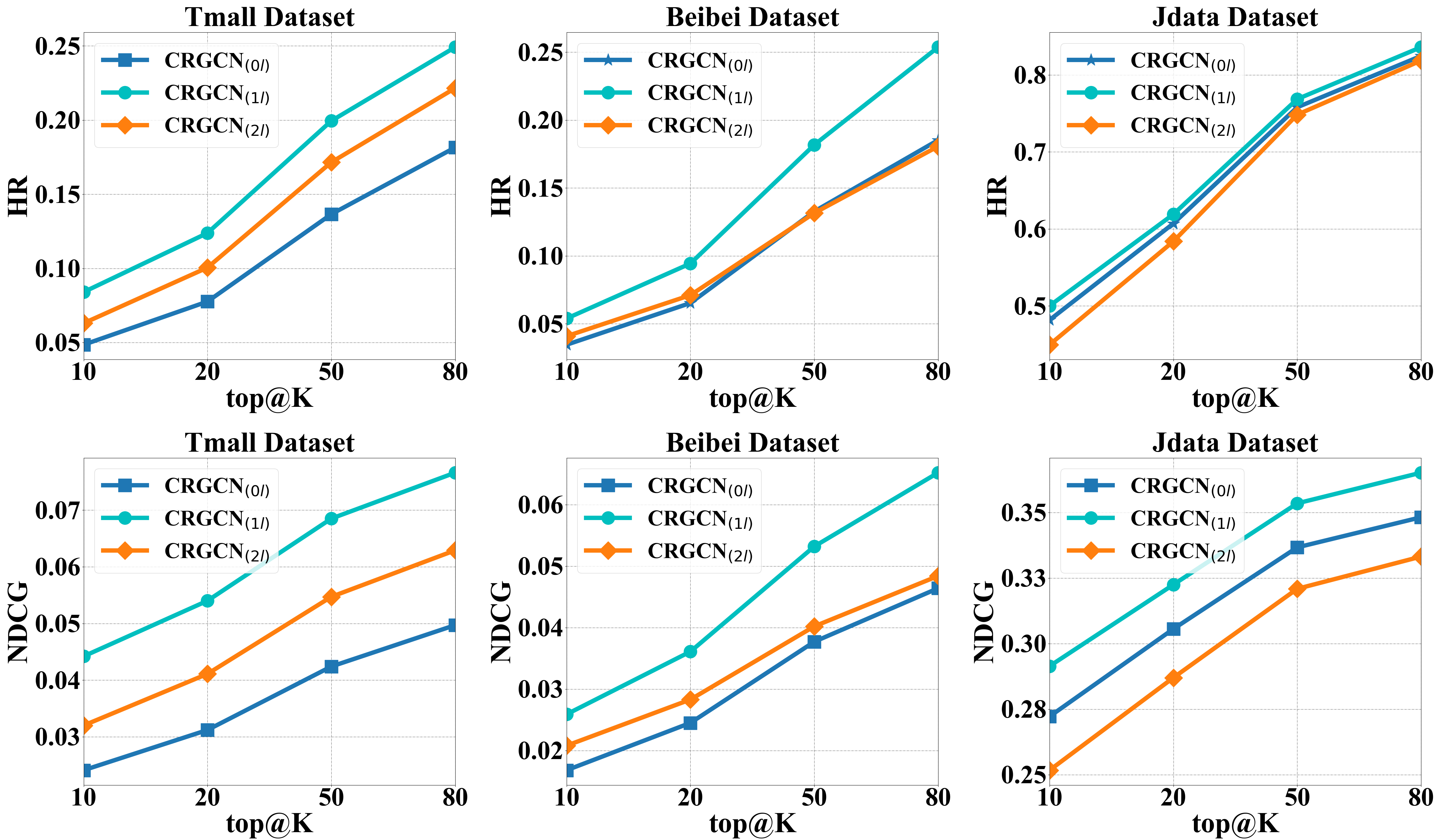}
  \caption{Effects of different GCN layers on the final performance of our model.}
  \Description{}
  \label{fig:gcn_layer}
\end{figure}

\subsubsection{\textbf{Effects of the number of GCN layers}} 
In this section, we study the effects of the number of GCN layers on the model performance. For simplicity, we use the same number of layers for different behaviors. The experimental results on Tmall, Beibei, and Jdata datasets are shown in Fig.~\ref{fig:gcn_layer}, in which CRGCN$_{(0l)}$, CRGCN$_{(1l)}$ (our model), and CRGCN$_{(2l)}$ represent GCN with 0, 1, and 2 layers, respectively. In particular, when the layer number of the GCN is 0, our CRGCN model degenerates into multi-task learning with shared embeddings.

By comparing the results of CRGCN$_{(0l)}$ and CRGCN$_{(1l)}$ in Fig.~\ref{fig:gcn_layer}, we find out that the use of GCN to learn user preferences from behaviors can greatly improve the model performance. Besides, the performance of the model decreases with the increase of GCN layers, which can be seen from the results of CRGCN$_{(1l)}$ and CRGCN$_{(2l)}$ on Tmall and Beibei datasets. In previous work like LightGCN~\cite{LightGCN} or NGCF~\cite{Wang0WFC19}, the best performance is often obtained when stacking 2 or 3 layers, as stacking more layers can exploit higher-order neighboring information to learn better user preferences. However, in our model, the best performance is achieved by using only one layer. This is because our model applies the GCN model to learn user preferences from multiple behaviors in a cascading way. The embedding learned from a GCN module (based on behavior) is passed to the next GCN for refinement. And we assume that the behavior that performs the latter can express more accurate user preferences. For example, the \emph{view} behavior only expresses a general interest of a user towards an item. It is not sure whether the user is really interested in an item yet before the user gets more details of the item. And the number of interactions based on the \emph{view} behavior is significantly larger than other behaviors (see Table~\ref{tab:dataset}). In other words, the items in the latter behaviors can better reflect a user's preference (\textit{i.e.}, \emph{buy}->\emph{collect}->\emph{view}). When stacking more layers on the graph of \emph{view} behavior, it may introduce noisy information into the learning process, which will negatively affect the subsequent embedding learning based on the latter behaviors, which can be confirmed in the results of Jdata dataset.  

To validate this viewpoint, we perform an additional experiment. In this experiment, we change the number of GCN layers for different behaviors, and each time we make the change for only one behavior. The behavior modeling order is \emph{view}->\emph{collect}->\emph{cart}->\emph{buy} for Tmall and Jdata datasets, and \emph{view}->\emph{cart}->\emph{buy} for Beibei dataset, respectively. For ease of presentation, we use a list to represent the number of GCN layers in the corresponding behavior. For example, in the Tmall dataset, $[2, 1, 1, 1]$ indicates that there are 2 GCN layers for \emph{view} and 1 GCN layer for all other behaviors. The same definition is used for Beibei and Jdata datasets. The experimental results are shown in Table~\ref{tab:higher_order}. Generally, the earlier the behavior of using more layers, the worse the performance.
In addition, with the use of more layers in the last behavior (\textit{i.e.}, \emph{buy} behavior), the performance can be further improved (see T$_5$ over T$_1$, B$_4$ over B$_1$, and J$_5$ over J$_1$). The best performance is achieved when using two GCN layers on the \emph{buy} behavior, which is also the target behavior. The results can well support our explanation for the results in Fig.~\ref{fig:gcn_layer}. In addition, this also verifies the benefits of using GCN to exploit higher-order neighbors in the graph to learn user preferences. However, it is better to confirm that the neighbors are indeed positively related. This is also one of the reasons that stacking more layers may cause performance degradation\footnote{Another reason is the over-smoothing problem, which is an inherent problem for GCN models.}. Because after a few hops, it is hard to differentiate the relevance of high-order neighbors.

\begin{table}[htb]
  \caption{Performance with different GCN layers for different behaviors.}
  \label{tab:higher_order}
  \resizebox{0.90\textwidth}{!}{
  \begin{tabular}{cccccccc}
  \toprule
  \textbf{Dataset} & \textbf{Evaluation metrics} & \multicolumn{6}{c}{\textbf{Statistic}} \\ 
  \hline
  \multirow{4}{*}{\textbf{Tmall}} & \multirow{2}{*}{\textbf{View-\textgreater Collect-\textgreater Cart-\textgreater Buy}} & $\mathbf T_1$ & $\mathbf T_2$ & $\mathbf T_3$ & $\mathbf T_4$ & $\mathbf T_5$ & $\mathbf T_6$ \\ \cmidrule(l){3-8} 
  & & {[}1, 1, 1, 1{]} & {[}2, 1, 1, 1{]} & {[}1, 2, 1, 1{]} & {[}1, 1, 2, 1{]} & {[}1, 1, 1, 2{]} & {[}1, 1, 1, 3{]} \\ \cmidrule(l){2-8}
  & \textbf{HR@20}   & \underline{0.1223} & 0.1036 & 0.1165 & 0.1211 & \textbf{0.1233} & {0.1204} \\
  & \textbf{NDCG@20} & \underline{0.0532} & 0.0440 & 0.0489 & 0.0528 & \textbf{0.0534} & {0.0514} \\
  \hline
  \multirow{4}{*}{\textbf{Beibei}} & \multirow{2}{*}{\textbf{View-\textgreater Cart-\textgreater Buy}} & $\mathbf B_1$  & $\mathbf B_2$ & $\mathbf B_3$ & $\mathbf B_4$ & $\mathbf B_5$ & - \\ \cmidrule(l){3-8} 
  & & {[}1, 1, 1{]}  & {[}2, 1, 1{]} & {[}1, 2, 1{]} & {[}1, 1, 2{]} & {[}1, 1, 3{]} & - \\ \cmidrule(l){2-8}
  & \textbf{HR@20}   & \underline{0.0812} & 0.0658 & 0.0598 & \textbf{0.0927} & 0.0778 & - \\  
  & \textbf{NDCG@20} & \underline{0.0312} & 0.0291 & 0.0248 & \textbf{0.0369} & 0.0290 & - \\
  \hline
  \multirow{4}{*}{\textbf{Jdata}} & \multirow{2}{*}{\textbf{View-\textgreater Collect-\textgreater Cart-\textgreater Buy}} & $\mathbf J_1$ & $\mathbf J_2$ & $\mathbf J_3$ & $\mathbf J_4$ & $\mathbf J_5$ & $\mathbf J_6$ \\ \cmidrule(l){3-8} 
  & & {[}1, 1, 1, 1{]} & {[}2, 1, 1, 1{]} & {[}1, 2, 1, 1{]} & {[}1, 1, 2, 1{]} & {[}1, 1, 1, 2{]} & {[}1, 1, 1, 3{]} \\ \cmidrule(l){2-8}
  & \textbf{HR@20}   & \underline{0.1223} & 0.1036 & 0.1165 & 0.1211 & \textbf{0.1233} & {0.1204} \\
  & \textbf{NDCG@20} & \underline{0.0532} & 0.0440 & 0.0489 & 0.0528 & \textbf{0.0534} & {0.0514} \\
  \bottomrule 
  \end{tabular}
  }
\end{table}

\begin{figure}[htb]
  \centering
  \includegraphics[width=\textwidth]{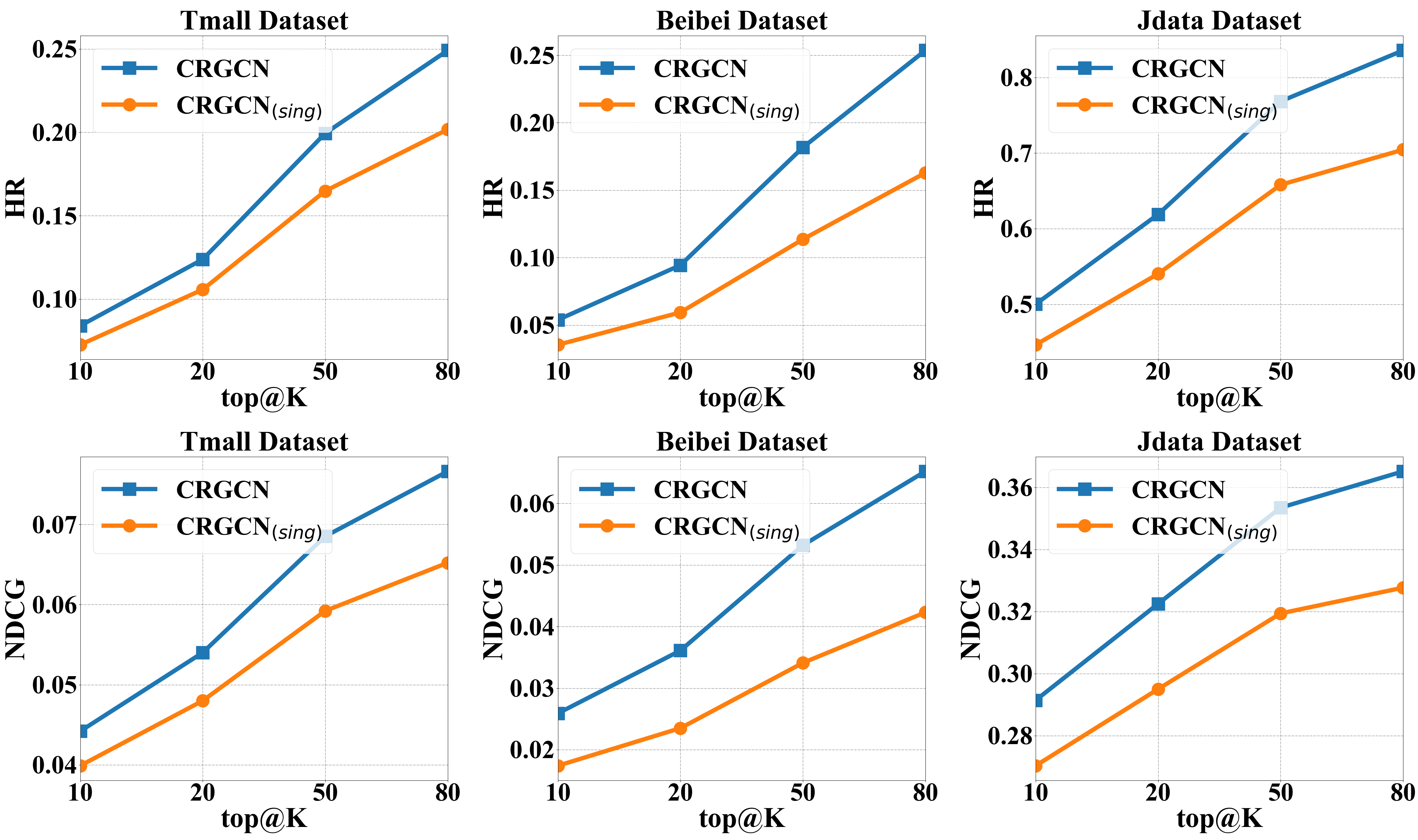}
  \caption{Performance comparisons between the multi-task learning and single-task learning.}
  \Description{}
  \label{fig:one_behavior}
\end{figure}

\subsubsection{\textbf{Effects of multi-task learning}}

To verify the effectiveness of multi-task learning in CRGCN, we compare CRGCN with the one using the prediction loss of the target behavior (\textit{i.e.}, \emph{buy} behavior) for training (denoted as CRGCN$_{(sing)}$) in experiments. All the other parts are kept the same as CRGCN. The experimental results on Tmall, Beibei, and Jdata datasets are shown in Fig.~\ref{fig:one_behavior}.

From the results, we can observe that the performance of CRGCN is consistently better than that of CRGCN$_{(sing)}$ in terms of both HR@K and NDCG@K metrics, indicating that multi-task learning can indeed significantly improve the prediction accuracy for the target behavior. The reason for this is that multi-task learning can jointly train the outputs of all residual blocks at the same time, making good use of each behavior to learn user preferences. This can not only learn better embeddings from the interaction data of each behavior, but also facilitate the continuous refinement of user preferences with a cascading structure by using the multi-task learning framework.

\subsubsection{\textbf{The importance of different weights}}

\begin{figure}[htb]
  \centering
  \includegraphics[width=\textwidth]{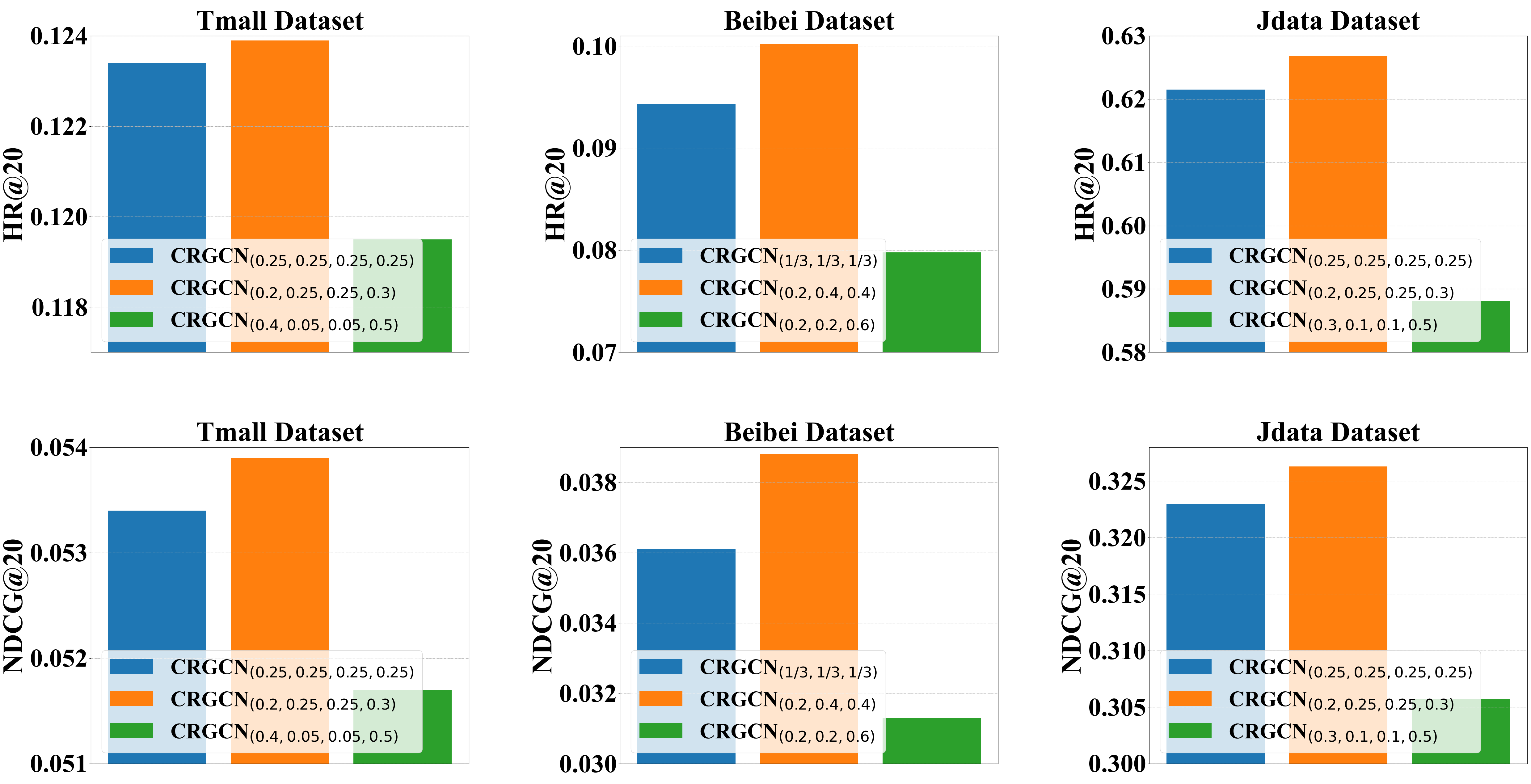}
  \caption{Performance of assigning different weights to different tasks in multi-task learning.}
  \Description{}
  \label{fig:weight_learning}
\end{figure}

In this work, we mainly focus on studying the potential of our model in exploiting the cascading effects of multi-behavior in embedding learning. Therefore, we treat different tasks equally in the loss function for simplicity. Intuitively, different behaviors may have different effects on the target behavior. Thus, we carry out experiments here to analyze the impact of loss function on different tasks with different weights. Specifically, we test different combinations of weight settings and we did not exhaust all the possibilities. The experiment results are reported in Fig.~\ref{fig:weight_learning}.

From the results, we can see that the equal setting indeed cannot achieve the best results. Better results can be achieved by assigning different weights to different tasks. In general, assigning higher weights to latter behaviors can yield relatively better performance, which is expected because the latter behaviors are closer to the target behaviors. But still, it needs to carefully tune the weights for the optimal performance. Because this is not our main focus in this study, and the simple equal setting can already achieve impressive improvement over the most recently proposed models, we did not carefully tune the weights of different tasks for the best performance. This experiment can already verify that the importance of different tasks should be different in the multi-task learning for better performance. We would like to leave the exploration of automatically learning the best setting in the future.

\subsection{\textbf{Influence of multi-behaviors (RQ3)}}
The underlying assumptions of our model include that different types of behaviors provide valuable information about user's preference and the order of behaviors (\textit{e.g.}, \emph{view}->\emph{cart}->\emph{buy}) matters. Specifically, the next behavior provides more specific information which can help us refine the user's preference. To validate the assumptions, in this section, we study the effects of multi-behavior information on the recommendation performance from two aspects: 1) the number of behaviors, and 2) the order of behaviors. 

Before reporting the experimental results, we would like to describe the behavior sequence on the Tmall, Beibei, and JD platforms. When shopping on Beibei, users must follow the order of \emph{<view, cart, buy>}, which is fixed. In contrast, on Tmall and JD, after the behavior of \emph{view}, users can perform the behaviors of \emph{collect} or \emph{cart} and then \emph{buy}, or directly go to the final behavior of \emph{buy}. Specifically, the possible behavior sequences can be \emph{<view, buy>}, \emph{<view, cart, buy>}, \emph{<view, collect, buy>}, \emph{<view, collect, cart, buy>}. Besides the above sequences, we also added other behavior sequences as competitors for analysis.

\begin{figure*}[htb]
  \centering
  \includegraphics[width=\textwidth]{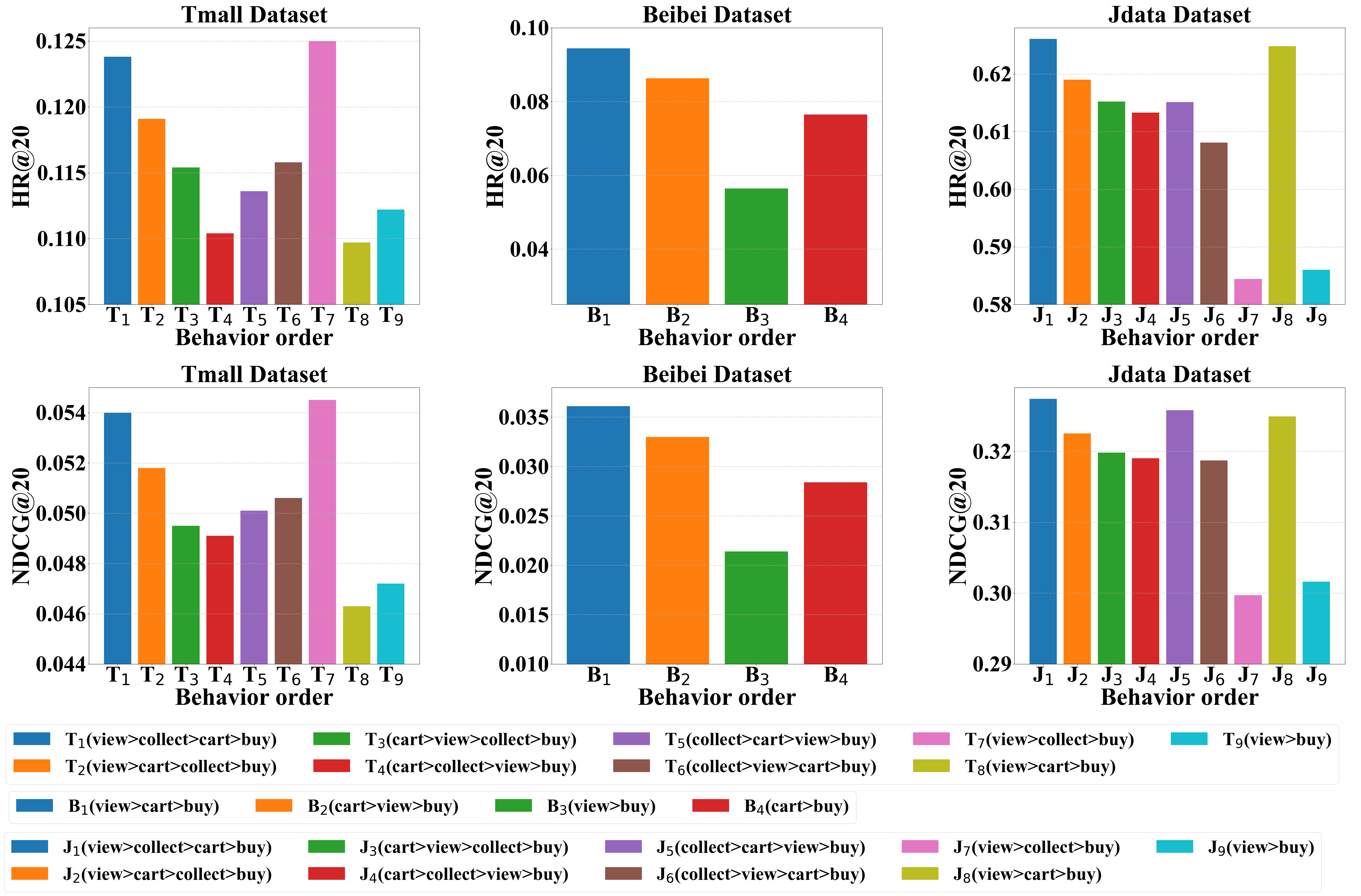}
  \caption{Performance comparison of CRGCN \textit{w.r.t.} different order and number of behaviors on three datasets.}
  \Description{}
  \label{fig:behavior_order}
\end{figure*}

The experimental results on Tmall, Beibei, and Jdata are shown in Fig.~\ref{fig:behavior_order}, respectively. From the results on Tmall and Jdata datasets, it is interesting to find that the increase of behavior numbers does not necessarily improve the performance and even may cause performance degradation, especially when the behaviors have not been taken into consideration in the correct order (\textit{i.e.}, the order of behaviors that users often perform in real scenarios). We first take a look at the four behavior orders that users often perform in real scenarios: T$_9$/J$_9$: \emph{view}->\emph{buy}, T$_7$/J$_7$: \emph{view}->\emph{collect}->\emph{buy}, T$_8$/J$_8$: \emph{view}->\emph{cart}->\emph{buy}, T$_1$/J$_1$: \emph{view}->\emph{collect}->\emph{cart}->\emph{buy}. The performance of J$_8$ is much better than that of J$_9$, which means that considering the \emph{cart} behavior helps to better model the user's shopping process. The performance of T$_8$ is worse than that of T$_9$ because the \emph{cart} behavior data is too sparse in Tmall. As a result, we cannot learn good representations of users and items for this behavior based on such sparse data, and it will hurt the embedding learning process when taking it into the sequence. The comparable performance of T$_1$ and T$_7$ can also validate this point. By contrast, the addition of \emph{collect} behavior significantly improves the performance (T$_7$ over T$_9$) for Tmall dataset, the reason is that the \emph{collect} behavior of Tmall dataset has more records (equivalent to the \emph{buy} behavior), so it can help to mine other aspects of user preferences.
For the performance with four behaviors in different orders, we can see that the more the behaviors are out of order (\textit{i.e.}, the farther away from the correct order), the worse the performance is. For example, when the \emph{view} behavior is the first behavior, the performance is better than that of other cases (see T$_1$/J$_1$ and T$_2$/J$_2$), and the performance is the worst when we put the \emph{view} behavior in the third place of the behavior sequence. This demonstrates our assumption that the next behavior in a sequence uncovers more information (than a previous behavior) about a user's preference, and our CRGCN can well model the cascading effects in multi-behaviors.




The performance on Beibei further validates the effectiveness of our model. Note that the behaviors must be taken in a fixed order on Beibei, \textit{e.g.}, \emph{view}->\emph{cart}->\emph{buy}. Obviously, with more behaviors in this order, we can infer the user's preference more accurately. The consistently better performance of B$_{1}$ over B$_{2}$ demonstrates that our model can well capture user preferences step by step based on the sequence of behaviors that are often taken by users in real scenarios. The reason that B$_{4}$ performs better than B$_{3}$ might be that: 1) the \emph{cart} behavior reveals more information about user preferences than of the \emph{view} behavior; 2) there are more overlapped preferences that encoded between the behaviors of \emph{cart} and \emph{buy} than the one that encoded between the behaviors of \emph{view} and \emph{buy}. In other words, the connection between \emph{buy} and \emph{cart} behavior is closer than that between the \emph{buy} and \emph{view} behavior. Our model is designed to capture the connections between behaviors and can make more use of the closer connections. 

\subsection{Performance on Cold-start Users (RQ4)}
The cold-start user problem is an inherently challenging problem in recommender systems. When the interactions between users and items become sparse, the performance of most collaborative filtering-based recommendation models decreases sharply. In real scenarios, the \emph{buy} behaviors of users are often very sparse, which severely limits the effectiveness of recommendation models. The advantage of multi-behavior recommendation models is that they can leverage the information of other behaviors to make a recommendation for the final \emph{buy} behavior, so as to alleviate the cold-start problem of users. In this section, we would like to testify the effectiveness of our model in tackling this problem.

\begin{figure}[htb]
  \centering
  \includegraphics[width=\textwidth]{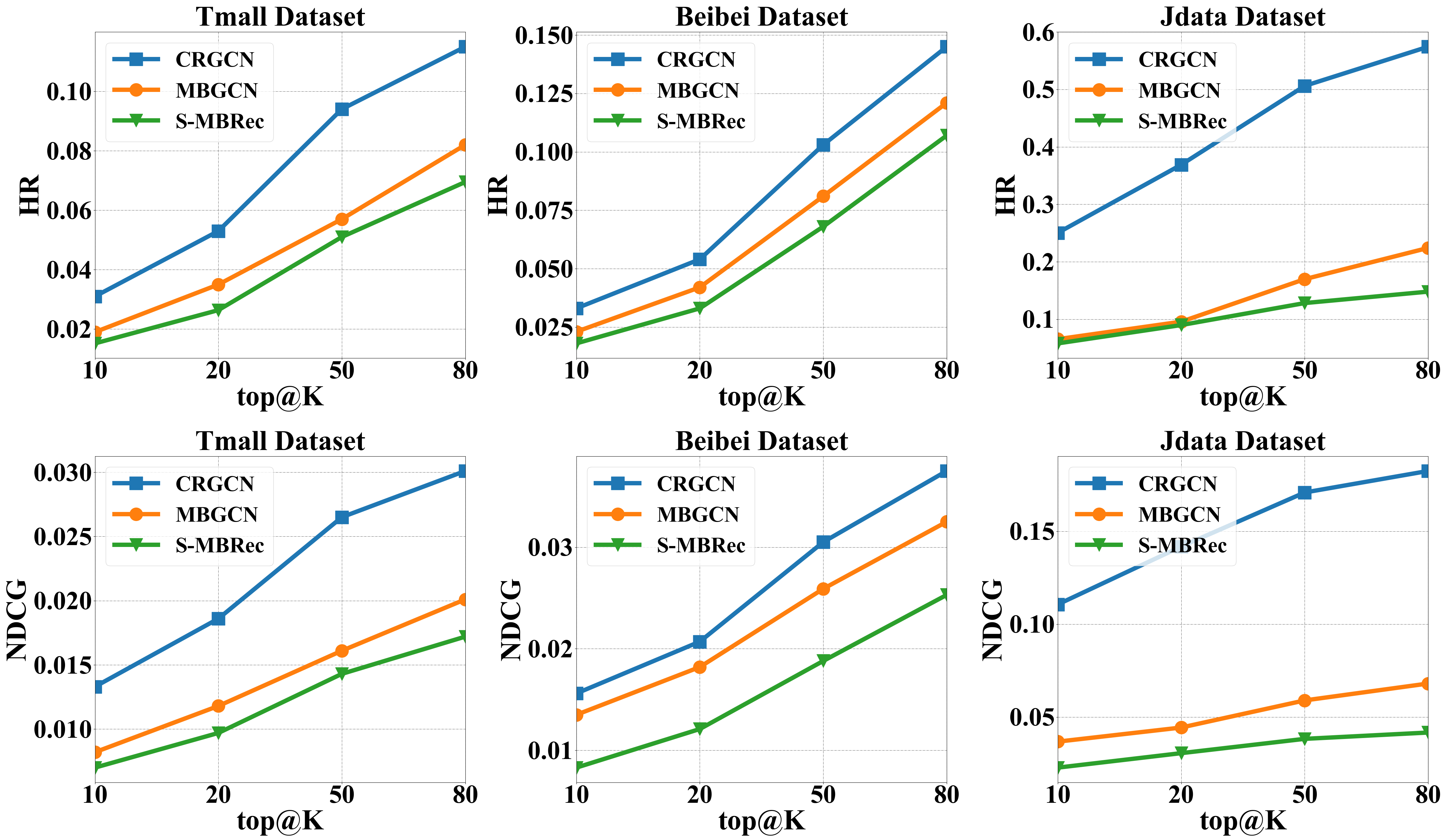}
  \caption{Performance comparisons among CRGCN, S-MBRec, and MBGCN for cold-start users.}
  \Description{}
  \label{fig:cold_start}
\end{figure}

We compare our CRGCN with two models S-MBRec and MBGCN, where S-MBRec is the best baseline, and MBGCN yields better performance over S-MBRec in dealing with the cold-start problem. 
To perform the study, we randomly select 1,000 users in the test set as cold-start users and remove their records of \emph{buy} behaviors from the training set. To be more specific, we remove all the \emph{buy} behavior interactions of these 1,000 users in the training set. In addition, for all the user-item pairs involved in the removed \emph{buy} behavior interactions, their interactions in all the other behaviors are also removed. For example, for a sampled user $u$, we not only remove her \emph{buy} behavior interaction with an item $i$, but all the other interaction behaviors with this item, such as \emph{view} and \emph{cart}, are also removed from the training data. This process is to ensure that the system does not have any prior information about the user's preference for the test item. For other behaviors, we delete user-item interaction pairs observed in our randomly selected collection. These 1,000 users are simulated as the hard cold-start users with no \emph{buy} behavior records. Then we exploit the remaining users' interaction records to train the model, and follow the same settings as described in section~\ref{Settings}. Finally, we use the trained model to make personalized recommendations for these 1,000 cold-start users.

The recommendation results of the three models are shown in Fig.~\ref{fig:cold_start} in terms of HR@K and NDCG@K. 
It can be observed that our CRGCN model consistently outperforms MBGCN and S-MBRec with a large margin. Compared with MBGCN, the average improvement of our model in terms of HR@K and NDCG@K are 103.86\% and 94.52\% on the Tmall dataset, 29.76\% and 15.61\% on Beibei dataset, and 230.82\% and 194.99\% on Jdata dataset. The superiority demonstrates that our model can better utilize the assistive behavior information to learn users' preferences for the target behavior recommendation. This should be attributed to the cascading residual design, which can effectively leverage the cascading behaviors to refine the user preference embedding. In addition, we find that although S-MBRec is the best baseline, but it does not perform as well as MBGCN in dealing with the cold-start problem. This is because MBGCN has an item-based scoring module to leverage the item-item relations, which can provide additional information for user preference modeling.

\subsection{Efficiency Analysis (RQ5)}
Another advantage of our CRGCN model is its light design without introducing additional weights (compared with the single-behavior models) as other multi-behavior recommendation models, which can greatly save the time cost in the training process. To evidently demonstrate the computing efficiency of the model, we compare CRGCN with several representative baselines in the same settings based on the average training time for one epoch and the number of epochs to converge. Besides, we also report the time it takes for each method to perform the same prediction. The results are shown in Table~\ref{tab:complexity}. All methods in this experiment are implemented by Pytorch, in which the NMTR-NCF model only considers two types of behaviors (\textit{i.e.}, \emph{view} and \emph{buy}) on Tmall and Jdata dataset. The environment settings are as follows: \textbf{CPU:} Intel(R) Xeon(R) CPU E5-2650 v4 @ 2.20GHz, \textbf{GPU:} GeForce RTX 2080 Ti Rev. A, \textbf{Batch size:} 1,024, \textbf{Embedding size:} 64.


\begin{table}[htb]
\caption{Comparison of computing efficiency for the training of each epoch.}
\label{tab:complexity}
\resizebox{\textwidth}{!}{
    \begin{tabular}{cccccccccc}
	\toprule
    \multirow{2}{*}{\textbf{Metric}} & \multirow{2}{*}{\textbf{Dataset}} & \multicolumn{2}{c}{\textbf{One-behavior}} & \multicolumn{6}{c}{\textbf{Multi-behavior}} \\ \cmidrule(l){3-4} \cmidrule(l){5-10}
                                      & & MF-BPR & LightGCN & R-GCN & NMTR-NCF & MBGCN & GNMR & S-MBRec & \textbf{CRGCN} \\ \hline
    \multirow{3}{*}{\textbf{Training(s)}} & \textbf{Tmall}  & 2.28 & 3.58 & 23.72 & 13.30 & 106.72 & 112.35 & 109.79 & \textbf{10.66}  \\
                                          & \textbf{Beibei} & 1.74 & 2.86 & 30.35 & 20.19 & 139.36 & 119.97 & 158.61 & \textbf{6.78}  \\
                                          & \textbf{Jdata}  & 5.14 & 7.92 & 21.89 & 23.76 & 105.69 & 120.00 & 168.29 & \textbf{19.58}  \\
    \bottomrule
    \end{tabular}
    }
\end{table}

From the results, we can see that CRGCN enjoys a good training efficiency compared with other multi-behavior models in terms of both the training time for each epoch and the required number of epochs for convergence. It is remarkable that our model is much more efficient than MBGCN and S-MBRec, which are the most competitive baselines in terms of the recommendation accuracy. Compared with the single-behavior LightGCN (only using the \emph{buy} behaviors), the total time cost is acceptable since our model uses much more interactions (see the number of interactions in Table~\ref{tab:dataset}). This is consistent with the complexity analysis in section~\ref{Complexity}, \textit{i.e.}, CRGCN has the same computing complexity as LightGCN. The time cost of R-GCN and NMTR-NCF is also much higher than our method, and their performance on accuracy falls far behind our model. Note that MTR-NCF only models two behaviors on the Tmall and Jdata datasets.  
In addition, it is worth mentioning that the training results of CRGCN (\textit{i.e.}, the final representation of users and items) can be conveniently saved, which means that our model has the same time consumption when making recommendation compared with MF-BPR, the most basic single-behavior recommendation model. The light design of CRGCN makes it more applicable for applications with large-scale datasets.

\section{conclusion} \label{conclusion}

In this work, we proposed a novel multi-behavior recommendation model named CRGCN, which exploits the cascading residual blocks to better mine user preferences expressed in a single behavior and the connection between different behaviors. 
Meanwhile, we designed the cascading residual structure to continuously refine user preferences and adopted the multi-task learning framework to optimize the model. Extensive experimental results on three real-world benchmark datasets demonstrate the superiority of our CRGCN model.
Further ablation studies verified the effectiveness of the components of CRGCN, including cascading residual blocks and multi-task learning. 
We also evaluated the performance of cold-start users and analyzed the model complexity, which confirms the high application value of our CRGCN in the real world.

In the future, we would like to explore how to combine and leverage micro-behavior (\textit{e.g.}, short-term multi-behavior interactions at the session level) and macro-behavior (\textit{e.g.}, long-term multi-behavior interactions) to further improve the performance of the personalized recommender systems.

\begin{acks}
  This work was supported in part by the National Natural Science Foundation of China under Grants 61902223, 62272254, and 61976042; in part by the Shandong Project towards the Integration of Education and Industry under Grants 2022PY009; in part by Young creative team in universities of Shandong Province under the grant 2020KJN012.
\end{acks}

\bibliographystyle{ACM-Reference-Format}
\bibliography{introduce}

\end{document}